\documentclass{article}

\usepackage{times}
\usepackage{graphicx} 
\usepackage{subfigure}

\usepackage{natbib}

\usepackage{algorithm}
\usepackage{algorithmic}
\usepackage[algo2e,linesnumbered,lined,boxed,vlined]{algorithm2e}

\usepackage{amsfonts}
\usepackage{amssymb}

\usepackage{hyperref}

\usepackage[accepted]{icml2016}

\usepackage{comment}
\usepackage{amsthm}
\usepackage{mathtools}
\usepackage{tabularx}
\usepackage{multirow}

\usepackage{lastpage}
\usepackage{fancyhdr}
\icmltitlerunning{Addressing A Mystery Using Data Visualization}

\begin{document}

\twocolumn[
\icmltitle{Addressing The Mystery of Population Decline of The Rose-Crested Blue Pipit In A Nature Preserve Using Data Visualization}

\icmlauthor{Benyamin Ghojogh}{bghojogh@uwaterloo.ca}
\icmladdress{Department of Electrical and Computer Engineering, 
\\Machine Learning Laboratory, University of Waterloo, Waterloo, ON, Canada}
\icmlauthor{Mark Crowley}{mcrowley@uwaterloo.ca}
\icmladdress{Department of Electrical and Computer Engineering, 
\\Machine Learning Laboratory, University of Waterloo, Waterloo, ON, Canada}
\icmlauthor{Fakhri Karray}{karray@uwaterloo.ca}
\icmladdress{Department of Electrical and Computer Engineering, 
\\Centre for Pattern Analysis and Machine Intelligence, University of Waterloo, Waterloo, ON, Canada}

\icmlkeywords{boring formatting information, machine learning, ICML}

\vskip 0.3in
]

\begin{abstract}
Two main methods for exploring patterns in data are data visualization and machine learning. The former relies on humans for investigating the patterns while the latter relies on machine learning algorithms. This paper tries to find the patterns using only data visualization. It addresses the mystery of population decline of a bird, named Rose-Crested Blue Pipit, in a hypothetical nature preserve. Different visualization techniques are used and the reasons of the problem are found and categorized. Finally, the solutions for preventing the future similar problems are suggested. This paper can be useful for getting introduced to some data visualization tools and techniques.
\end{abstract}

\section{Introduction}

\subsection{The Addressed Problem}

This paper addresses the VAST (Visual Analytics Science and Technology) 2017 challenge \cite{web_vast_challenge}. This challenge is about the mystery of population decline of a specific type of bird, named Rose-Crested Blue Pipit, in a hypothetical nature preserve. This preserve is close to a hypothetical city named Mistford. In this project, we aim to find out the possible reasons of this population decline using the provided dataset. This problem is sourced from various reasons which we address them one by one using pattern exploration by data visualization.
This paper can be useful for the reader to get introduced to some important data visualization tools and techniques. The visualizations in this paper are done in the R programming language. 

\subsection{Dataset}\label{section_dataset}

The dataset provided in \cite{web_vast_challenge} includes several data subsets. In this section, we briefly introduce the dataset and its data subsets; however, we defer the thorough explanation of the data to the relevant sections to analyze that piece of data. 

\begin{figure*}[!t]
\centering
\includegraphics[width=4.5in]{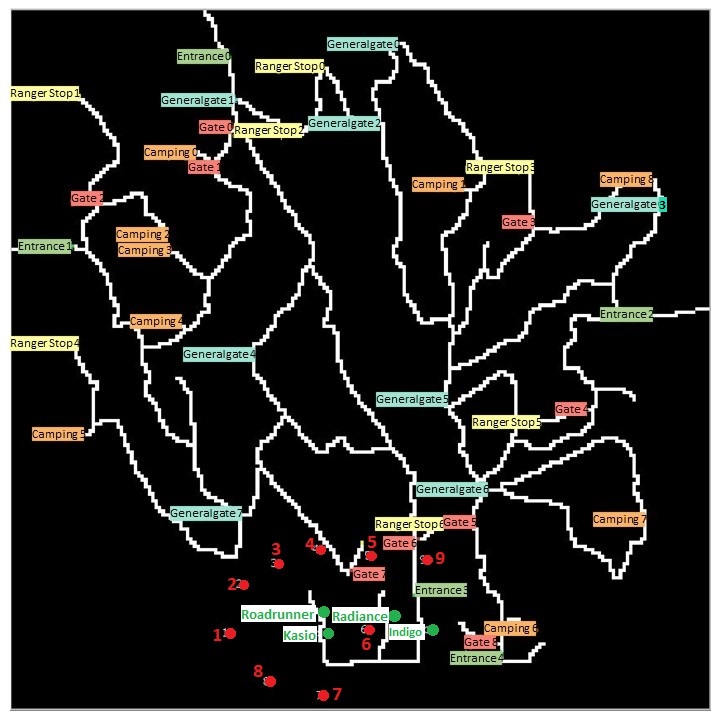}
\caption{The map of the nature preserve. The entrances, general gates, gates, ranger stops, and camping locations are identified by light green, light blue, light red, yellow, and orange colors. The nine chemical sensors are shown in red color and the factories at the south of the preserve are highlighted with dark green color.}
\label{figure_map}
\end{figure*}

\begin{itemize}
\item \textbf{Data subset 1:} This subset of data is provided for the mini-challenge 1 in the VAST challenge. A ($200\times 200$)-pixel map of the preserve is provided where five different color-coded types of gates in the preserve are shown in it. The five types of gates are entrances, general gates, gates, ranger stops, and camping. Moreover, the records of passing through these gates are reported in a file. These records are for seven different types of vehicles, i.e., two-axle car (or motorcycle), two-axle truck, three-axle truck, four-axle (and above) truck, two-axle bus, three-axle bus, and two-axle ranger truck. $171,477$ records are provided overall, each of which has information of time/date, car ID, car type, and gate name.
\item \textbf{Data subset 2:} This subset of data is provided for the mini-challenge 2 in the VAST challenge. The recorded data from nine different sensors, whose coordinates are shown in a provided map, are available. These sensors have collected samples of four specific chemicals emitted to the air by four factories close to the preserve. The coordinates of factories are also provided. $79,244$ records are provided overall. Moreover, meteorological data are also provided which show the date, wind direction, and wind speed. There are $708$ meteorological records available.
\item \textbf{Data subset 3:} This subset of data is provided for the mini-challenge 3 in the VAST challenge. Twelve multi-spectral six-channel images are provided which are taken from the preserve in different seasons of past few years. The channels of the images are blue, green, red, Near Infrared (NIR), Short-Wave Infrared (SWIR) 1, and Short-Wave Infrared (SWIR) 2. The size of every image is $650\times 650$ pixels. Moreover, a map of Boonsong Lake, a lake within the preserve (which is a hypothetical lake), is provided to help us find out the scale and orientation of the multi-spectral images.
\end{itemize}

As can be seen in the above explanations, this dataset includes different types of features, including numerical features, categorical features, time series, images, maps, etc, making it suitable for data visualization and pattern exploration. 

\section{Analysis of Vehicle Activities}

The data of vehicle activities and traffic through the preserve are given in order to investigate the possible unusual traffic patterns responsible for the decline of the Blue Pipit. 

\subsection{The Given Data}

The data subset 1 provides the information of the map of preserve where five different gates are shown on it. Moreover, the data of traffic of the vehicles of various types are given for analysis.

\begin{table}[!t]
\setlength\extrarowheight{5pt}
\centering
\scalebox{0.7}{    
\begin{tabular}{l | l | l}
\hline
\hline
\textbf{Gate} & \textbf{Allowed Vehicle} & \textbf{Role}\\
\hline
\hline
Entrance & All vehicles & Entring and leaving the preserve\\
\hline
General Gate & All vehicles & Recording flow of traffic\\
\hline
Gate & Rangers & Ranger activities beyond roadways\\
\hline
Ranger stop & All vehicles & Representing working areas for rangers\\
\hline
Camping & All vehicles & Recording visitors\\
\hline
Ranger base & Rangers & Rangers stay there when not working\\
\hline
\hline
\end{tabular}%
}
\caption{The different gates in the preserve.}
\label{table_gates}
\end{table}

\begin{figure}[!t]
\centering
\includegraphics[width=3.25in]{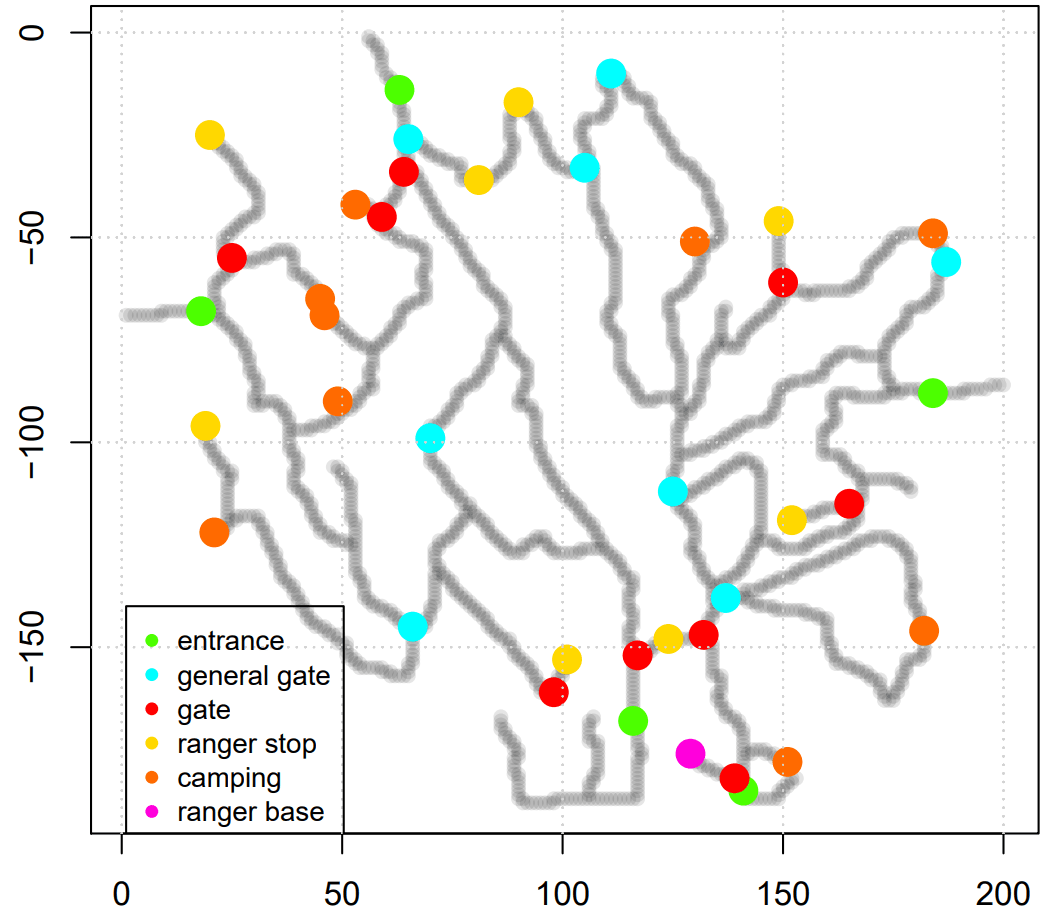}
\caption{The map and location of gates and roads in the nature preserve.}
\label{figure_plotted_map}
\end{figure}

\subsubsection{The Map of Nature Preserve}

The map of the nature preserve is shown in Fig. \ref{figure_map}. As can be seen in this map, there are five different gates named entrances, general gates, gates, ranger stops, and camping. When the vehicles pass these gates, the date and information of the passage are recorded. There is one additional gate, the ranger base gate, where the preserve rangers stay at when they are not working. Table \ref{table_gates} lists the existing gates in the preserve as well as explaining their roles. As this table shows, merely rangers are allowed to pass through the ``gates'' and the ranger base.

A ($200 \times 200$)-pixel bitmap image of the map is also given by the dataset. The gates are shown by colored pixels. Different gates are color-coded. The roads also have a different color. We read this bitmap image and by discriminating the colors, we find out the coordinates of the roads and different gates. Note that in this image, the $(0,0)$ pixel is the south-west corner while in real map, $(0,0)$ is at the north-west corner. Therefore, some manipulation is done for correcting the coordinates. The obtained map is shown in Fig. \ref{figure_plotted_map}. The gates are also color-coded in this map. 

\begin{figure}[!t]
\centering
\includegraphics[width=3in]{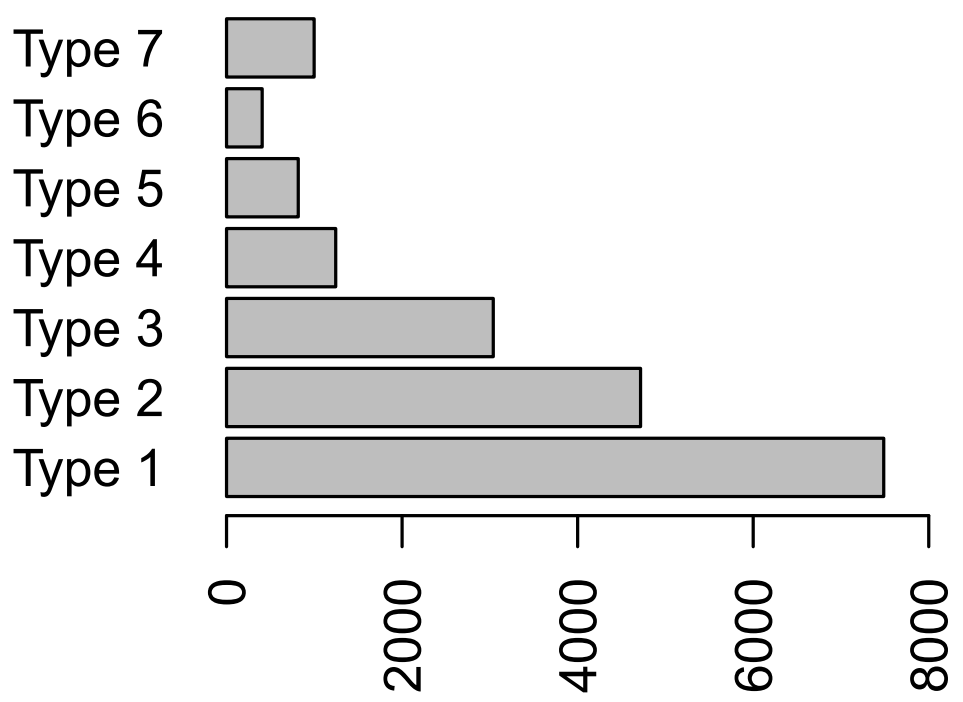}
\caption{The number of vehicles in different vehicle types.}
\label{figure_barplot_numberOfCars_inTypes}
\end{figure}

\begin{figure}[!t]
\centering
\includegraphics[width=3in]{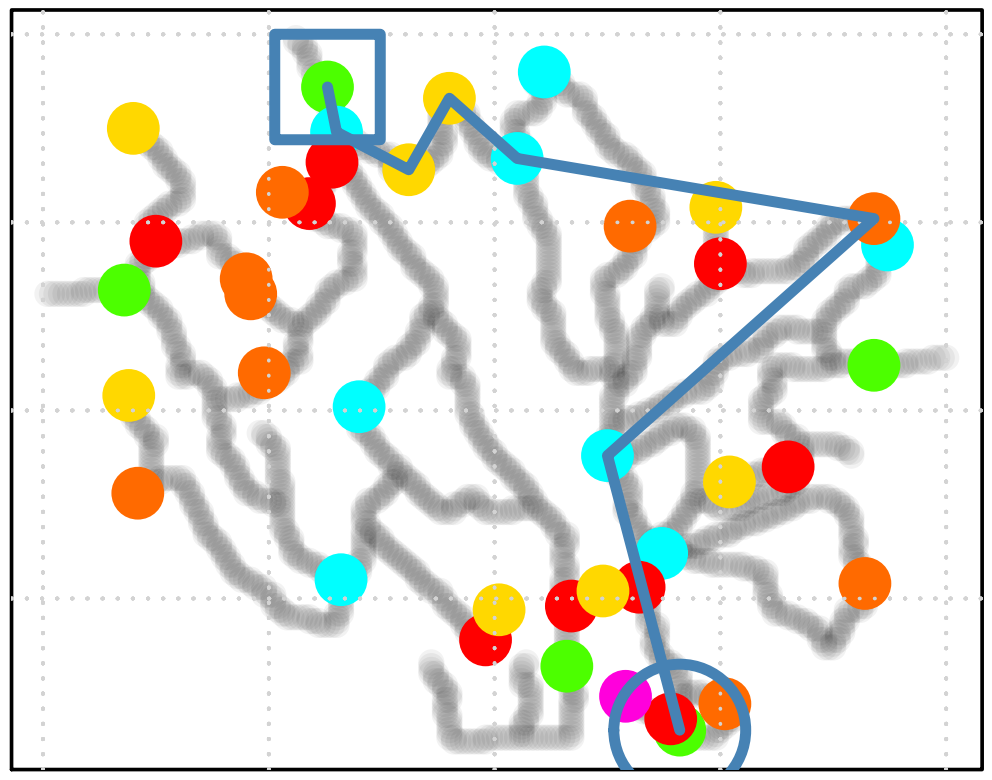}
\caption{The trajectory of a sample type-1 vehicle. The starting and ending points are shown by a circle and a square, respectively.}
\label{figure_trajectory_sampleCar}
\end{figure}

\subsubsection{The Traffic of Different Vehicles}

The recorded data for the traffic are categorized by the type of vehicle. Seven types of vehicles are type 1 (two-axle car or motorcycle), type 2 (two-axle truck), type 3 (three-axle truck), type 4 (four-axle and above truck), type 5 (two-axle bus), type 6 (three-axle bus), and type 7 (two-axle truck for rangers). Figure \ref{figure_barplot_numberOfCars_inTypes} shows the number of recorded vehicles in the different types. Type 1 has the most cardinality because most of regular visitors use this type of vehicle. Other types are mostly for construction or surveillance. 

The records of sensors are available in the dataset reporting the time, car ID, car type, and the gate for different passages during various time periods. For every specific car ID, we extract the data of that car; therefore, for every car a time series is obtained having the time of passing and the passed gate. If we replace the gate names by their 2D coordinates, this time series can be shown as a trajectory on the map. the trajectory of an example car is shown in Fig. \ref{figure_trajectory_sampleCar}. The trajectory connects the passed gates together in the order they have been passed. In the display of trajectory, we use a circle and a square for indicating the start and end points of the traversed path, respectively. This display gives us valuable information about the gates and the regions which are visited by the vehicle. 

\subsection{Analysis of Vehicle Traffic Based on Vehicle Type}

\subsubsection{Analysis of Vehicle Trajectories Based on Vehicle Type}

\begin{figure*}[!t]
\centering
\includegraphics[width=5in]{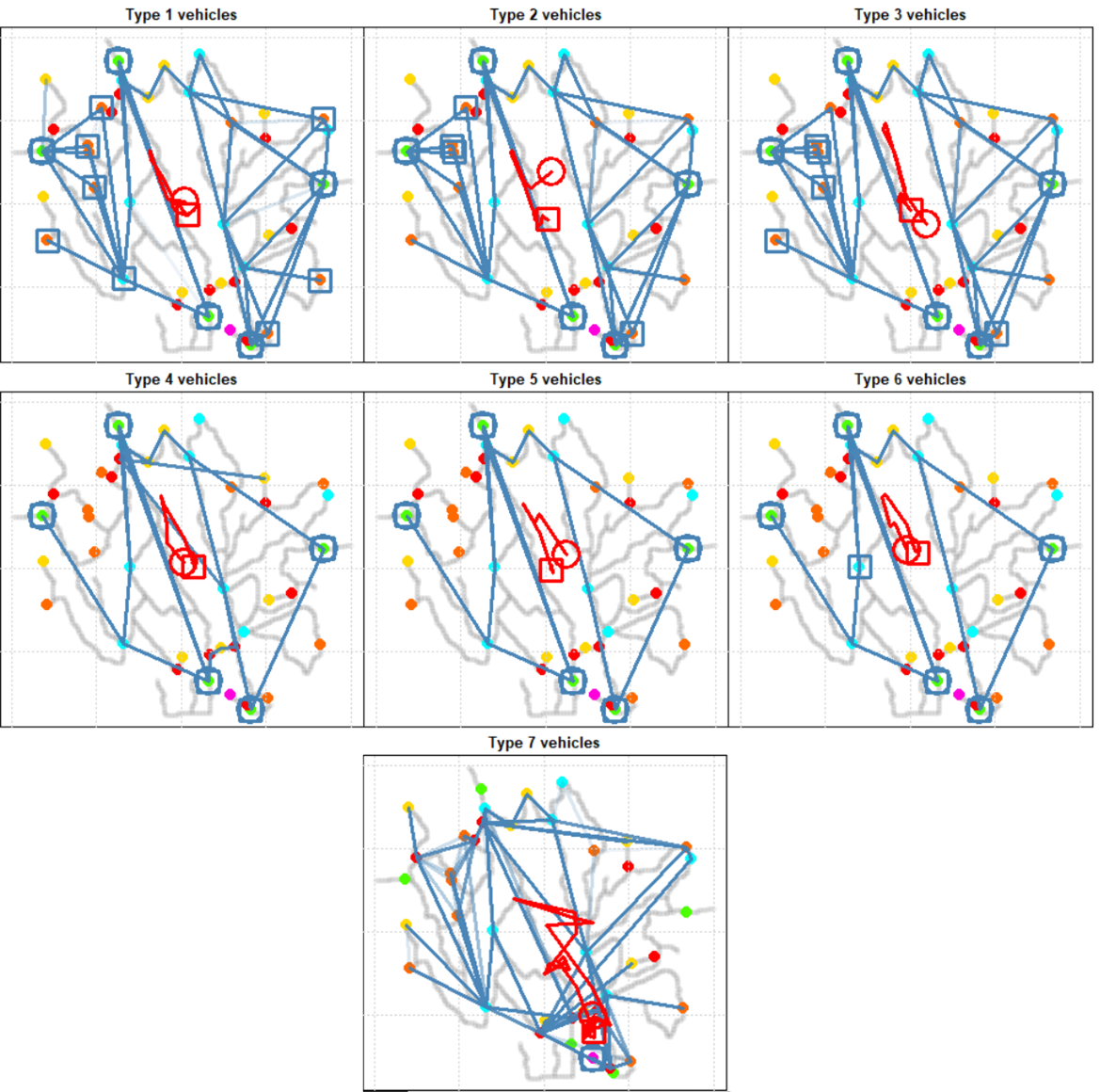}
\caption{The trajectories of vehicles in different types. The trajectory of every vehicle is shown by transparent blue color so that patterns and repeats can be obvious by bold blue color. The red trajectory is the mean trajectory of the cars in that type.}
\label{figure_trajectories_cars_of_types}
\end{figure*}

It is expected that most of the cars in the same type behave similarly and traverse similar paths. Therefore, for the cars of the same type, we plot the trajectory of the vehicles where the trajectories are transparent (alpha-blended) so that the trajectories which are repeated a lot get bold. In this way, the patterns of trajectories become obvious. The starting and end points are also shown so that we understand where the cars of a type have started and ended their trajectories. These trajectories are shown in Fig. \ref{figure_trajectories_cars_of_types}.

In the following, we analyze the trajectories of every vehicle type (the types of vehicles are determined by the dataset as mentioned in Section \ref{section_dataset}):
\begin{itemize}
\item \textbf{Type 1:} 
\begin{itemize}
\item Entering and leaving: Most of vehicles enter and leave from the entrance gates which is expected. However, \textbf{some of the trajectories end at the camping gates or a general gate}. This is unusual meaning that not any more passing is recorded after they have camped. This shows that some suspicious activities might have been done there.
\item Middle passed gates: The vehicles have passed through camping and general gates. This is expected because most of visitors, which come for camping, use type-1 vehicles. The middle ranger stops are also visited which is fine; however, \textbf{there is one trajectory passing the ranger stop at the top-left corner. This ranger stop is at the end of road and is usually passes by rangers.} Hence, this activity is also suspicious.
\end{itemize}

\item \textbf{Type 2:}
\begin{itemize}
\item Entering and leaving: Most of vehicles enter and leave from the entrance gates which is expected. However, \textbf{some of the trajectories end at some of the camping gates}. This is unusual and suspicious meaning that not any more passing is recorded after they have camped.
\item Middle passed gates: The vehicles have passed through camping and general gates which is fine.
\end{itemize}

\item \textbf{Type 3:}
\begin{itemize}
\item Entering and leaving: Most of vehicles enter and leave from the entrance gates which is expected. However, \textbf{some of the trajectories end at some of the camping gates}. This is unusual and suspicious meaning that not any more passing is recorded after they have camped.
\item Middle passed gates: The vehicles have passed through camping and general gates which is fine.
\end{itemize}

\item \textbf{Type 4:}
\begin{itemize}
\item Entering and leaving: All vehicles enter and leave from the entrance gates which is expected.
\item Middle passed gates: The vehicles have not passed through camping gates which is expected because they are heavy vehicles used for construction, etc. Passing through general gates is also fine. \textbf{However, some vehicles have passed through gates which are only allowed to be passed by the rangers. This definitely is illegal and suspicious. Moreover, some trajectories have visited the ranger stop at the end of road located at the up-right corner. This ranger stop is supposed to be passed mostly by rangers because of its location. This is also suspicious.}
\end{itemize}

\item \textbf{Type 5:}
\begin{itemize}
\item Entering and leaving: Most of vehicles enter and leave from the entrance gates which is expected. \textbf{However, some of the vehicles have ended their trajectories at general gate in the middle of the preserve. This is completely unusual and suspicious.}
\item Middle passed gates: The vehicles have passed through general gates and middle ranger stops which is fine.
\end{itemize}

\item \textbf{Type 6:}
\begin{itemize}
\item Entering and leaving: All vehicles enter and leave from the entrance gates which is expected.
\item Middle passed gates: The vehicles have passed through general gates and middle ranger stops which is fine.
\end{itemize}

\item \textbf{Type 7:}
\begin{itemize}
\item Entering and leaving: All vehicles enter and leave from the ranger base which is expected because vehicles of this type are all rangers.
\item Middle passed gates: The rangers are allowed to pass all gates. \textbf{However, not any ranger has visited and inspect two gates and a general gate in the east of the preserve. This gives some opportunity for the abusers to do some suspicious activities in the east of the preserve around those areas. The fact that they have missed those areas in their inspections is also suspicious.}
\end{itemize}
\end{itemize}

Moreover, the mean trajectory of the vehicles in a type might have interesting information. However, there is a problem that the lengths of trajectories are not necessarily equal. Therefore, we first find the longest trajectory of the vehicle type and then pad every trajectory of that type by repeating the last point of the trajectory at the end of it in order to make all trajectories of a type equal in length. Thereafter, we plot the mean trajectory on the map in order to show the overall behaviour of the vehicles of a type (see Fig. \ref{figure_trajectories_cars_of_types}).

We can see that the mean and the bold trajectories of the types 1, 2, and 3 are similar; and the trajectories of the types 4, 5, and 6 are similar. This makes sense because vehicles of types 1, 2, and 3 have less number of axles and the types 4, 5, and 6 are heavy vehicles with more number of axles. Also, we can see that the type-7 vehicles are visiting more places of the preserve which makes sense because they are supposed to inspect in the preserve.

\subsubsection{Analysis of Traffic Time Based on Vehicle type}

\begin{figure*}[!t]
\centering
\includegraphics[width=5.2in]{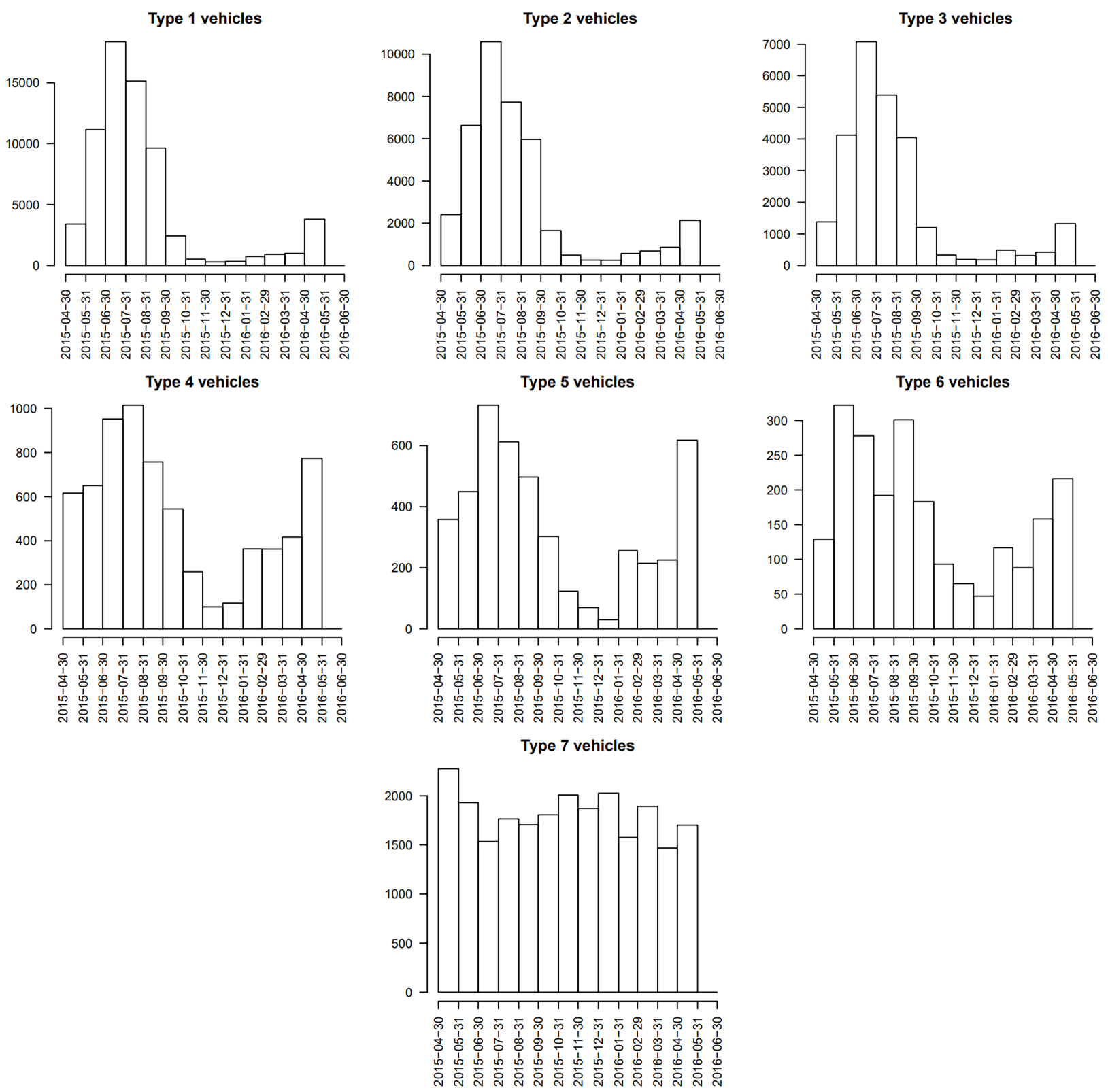}
\caption{The histogram of the dates when the vehicles of different types have passed through the preserve.}
\label{figure_hist_dates_carsInTypes}
\end{figure*}

In order to analyze the data of traffic in terms of time, we plot the histogram of the dates when the vehicles of different types have passed through the preserve. These histograms for different types of vehicles are shown in Fig. \ref{figure_hist_dates_carsInTypes}.

In the following, we list the analysis of the histograms:
\begin{itemize}
\item The histogram of types 1, 2, and 3 are similar, whereas the patterns of histograms of types 4, 5, and 6 are similar. This makes sense because vehicles of types 1, 2, and 3 have less number of axles and the types 4, 5, and 6 are heavy vehicles with more number of axles.
\item As expected, the rangers (type 7) have visited the preserve regularly in different times because they are supposed to inspect in the preserve in a regular schedule.
\item The number of vehicle records in the first types is much more than the number of last types of cars. This makes sense because the preserve is visited more by ordinary visitors than the construction trucks.
\item \textbf{Around June 2015, the traffic of all types of vehicles has increased a lot. This might be destructive to the nature preserve or might have had bad effects on the quality of life of flora and fauna in the preserve.}
\end{itemize}

\subsection{Analysis of Vehicle Traffic Based on Traffic Clusters}

Another way of analyzing the data is to cluster the traffic and then analyze the behaviours of the clusters rather than comparing the types of vehicles. 

\begin{figure}[!t]
\centering
\includegraphics[width=3.25in]{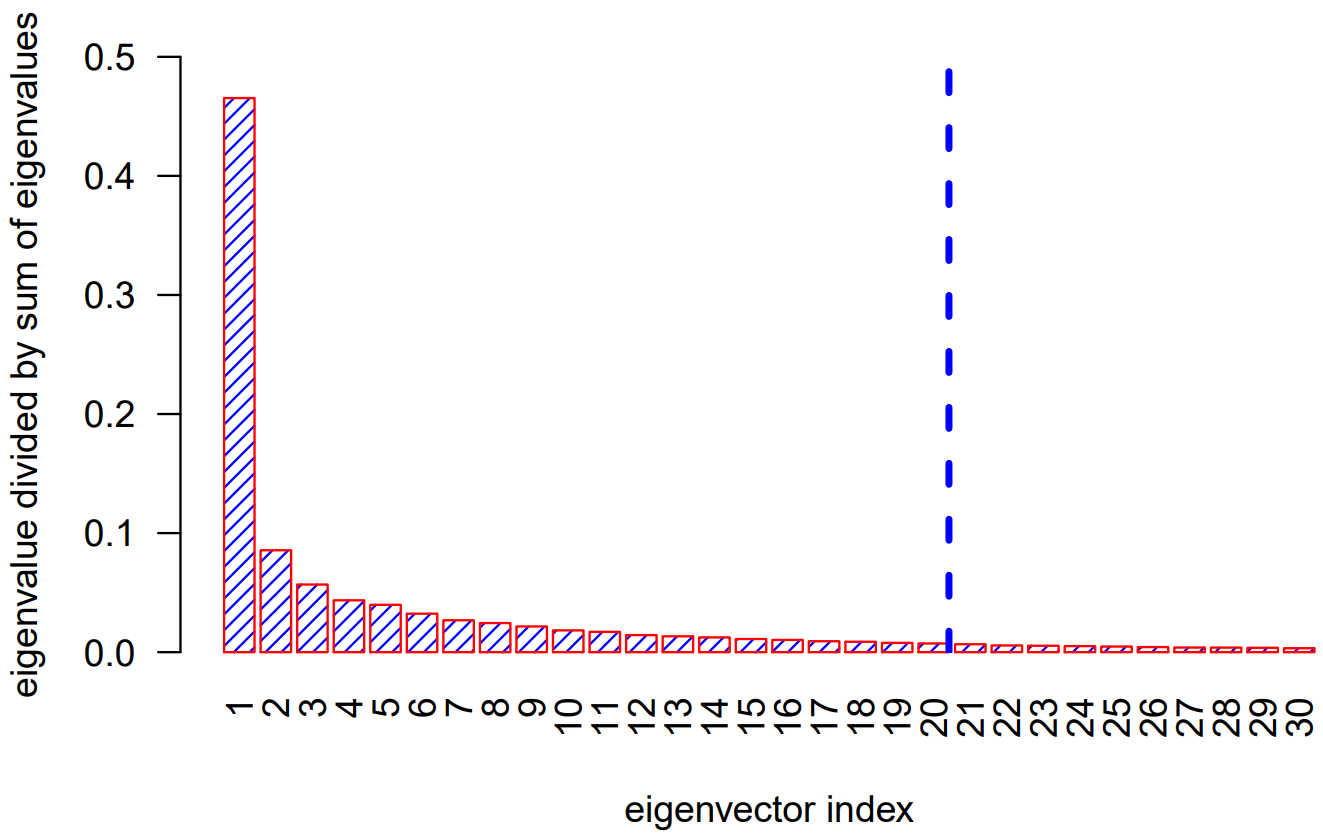}
\caption{Scree plot for determining the number of principal axes to choose.}
\label{figure_scree_plot}
\end{figure}

\subsubsection{Clustering the Traffic}

As the given data are huge and its processing takes a significant amount of time, we sample randomly from the data in order to have a subset of data. For a fair sampling, we randomly choose $100$ vehicles from every type and extract the data and time series of those cars from the dataset. This sampling is \textbf{stratified sampling} where the strata are the types of vehicles. It is known that stratified sampling is better than simple random sampling in terms of variance of estimation \cite{barnett1974elements}.

\begin{figure*}[!t]
\centering
\includegraphics[width=5.2in]{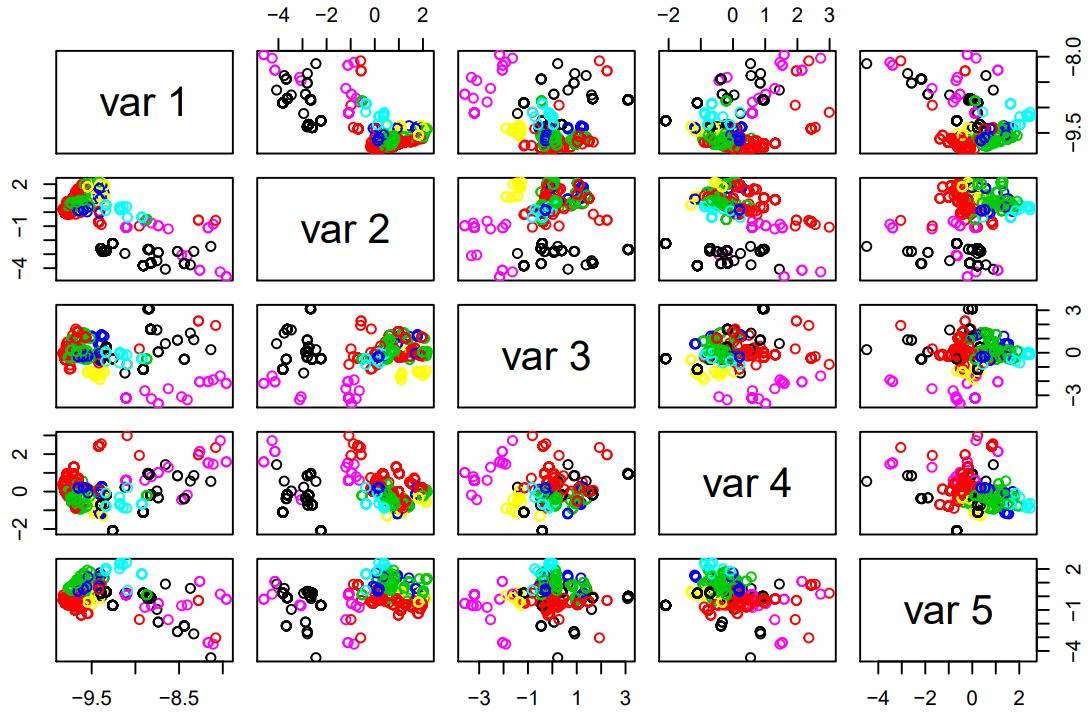}
\caption{Pairs plot of the first five dimensions of the data projected onto PCA subspace. The data of different clusters are color-coded.}
\label{figure_pairs_plot_clusters_of_projectedData}
\end{figure*}

After sampling from the dataset, we make the lengths of trajectories of the sampled vehicles equal by padding the last visiting point to the end of short trajectories. Then, we concatenate the $x$ and $y$ coordinates of the visited gates to make the trajectory of every vehicle as a vector. Putting the vectors of trajectories next to each other gives us a matrix.

Afterwards, we apply Principal Component Analysis (PCA) \cite{friedman2001elements} on this matrix-form data. The scree plot of the eigenvalues, which shows $\frac{\lambda_i}{\sum_j \lambda_j}$, is shown in Fig. \ref{figure_scree_plot}. As can be seen, the first $20$ eigenvectors seem to carry most of the information of variation of data. Therefore, we take the first $20$ principal axes and after projecting data onto these principal axes, we obtain the $20$-dimensional projected trajectories.

After applying PCA on data, we cluster the trajectories using K-means method \cite{friedman2001elements}. The number of clusters is reasonable to be $7$ because of seven different number of types of vehicles. The pairs plot (scatter plot matrix) of the first five dimensions of the projected data onto PCA subspace is shown in Fig. \ref{figure_pairs_plot_clusters_of_projectedData}. Every point in a scatter plot is the trajectory of a car. The points in this figure are color-coded according to their assigned clusters. As can be seen in this pairs plot, the clusters contain trajectories with different characteristics and values of dimensions. Hence, the clusters are properly dividing the trajectories into several groups.

\begin{figure}[!t]
\centering
\includegraphics[width=3in]{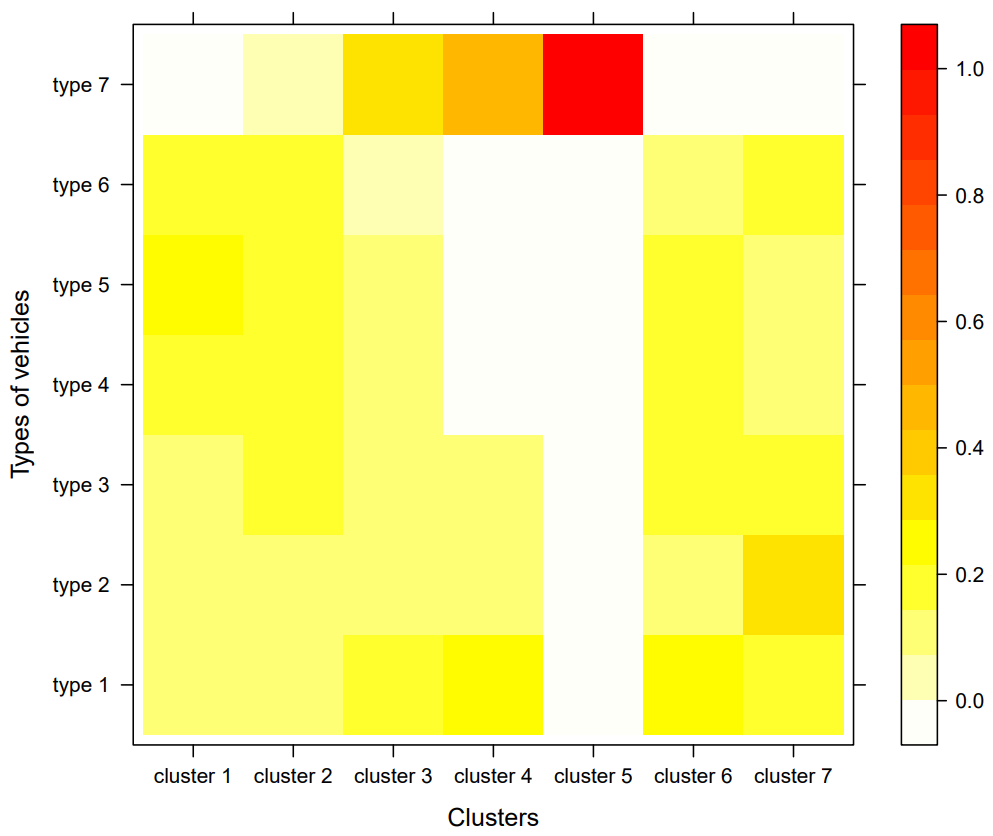}
\caption{Heat map of the proportion of cars of every type in different clusters.}
\label{figure_heatmap_n_cars_cluster_vs_types}
\end{figure}

In order to see the proportion of the number of vehicles of different types in every cluster, we plot the heat map of cardinality of each cluster in Fig. \ref{figure_heatmap_n_cars_cluster_vs_types}. As this figure shows, the rangers are mostly in one of the clusters, i.e., cluster 5. This makes sense because the behaviour of rangers, who should inspect in the preserve, is very different from other vehicles. Other types of vehicles are almost spread uniformly in the clusters because of close behaviour. This does not hurt our analysis because even some small variations between these clusters may have important information for us to discover.

\begin{figure*}[!t]
\centering
\includegraphics[width=5in]{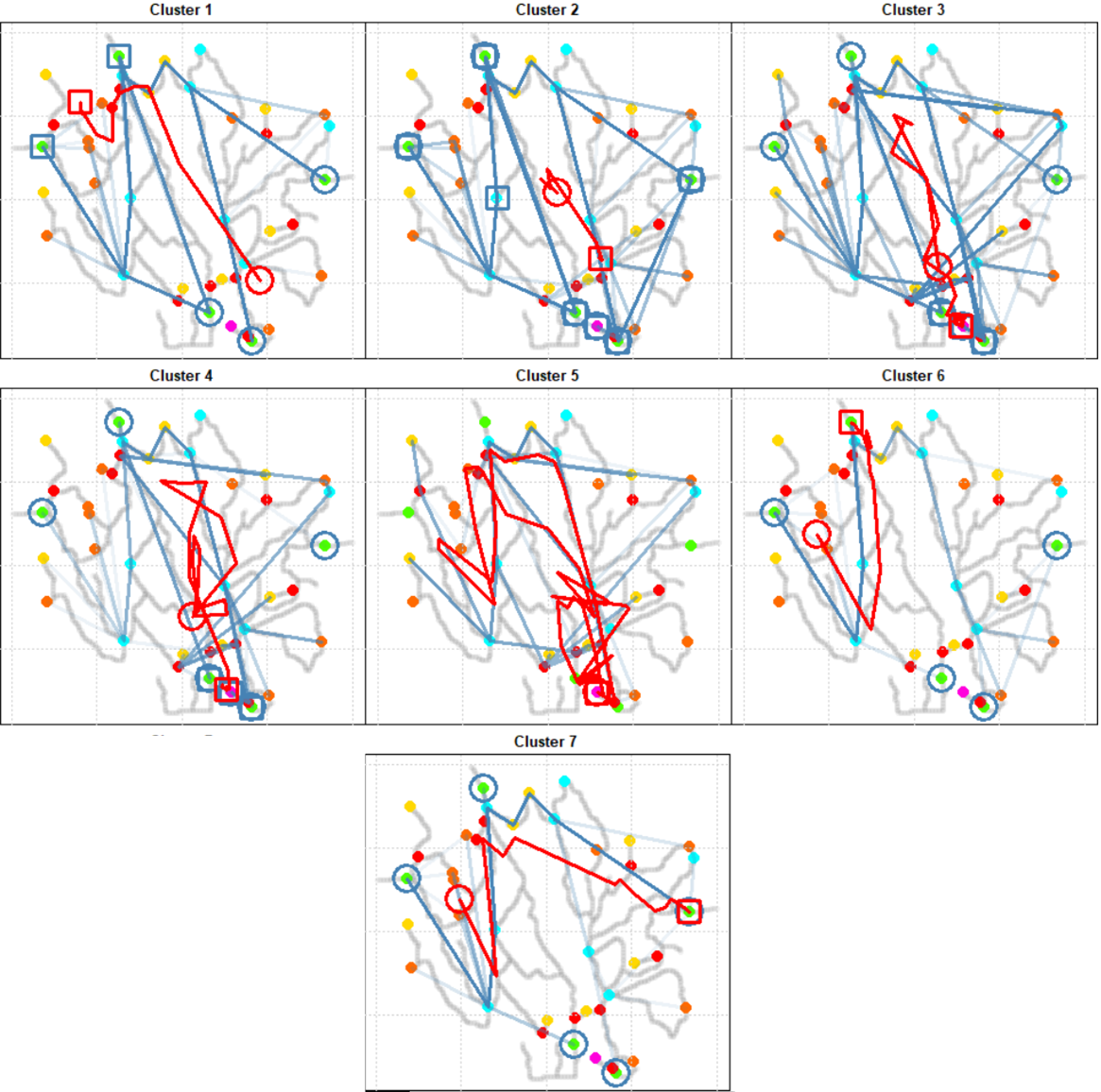}
\caption{The trajectories of vehicles in different clusters. The trajectory of every vehicle is shown by transparent blue color so that patterns and repeats can be obvious by bold blue color. The red trajectory is the mean trajectory of the cars in that cluster.}
\label{figure_trajectories_cars_of_clusters}
\end{figure*}

\subsubsection{Analysis of Vehicle Trajectories Based on Traffic Clusters}

For the cars of the same cluster, we plot the trajectory of the vehicles where the trajectories are transparent (alpha-blended) so that the trajectories which are repeated a lot get bold and the patterns show off. The starting and end points are also shown so that we understand where the cars of a type have started and ended their trajectories. These trajectories are shown in Fig. \ref{figure_trajectories_cars_of_clusters}. The mean trajectories of clusters are also shown in this figure.

\begin{figure*}[!t]
\centering
\includegraphics[width=5.5in]{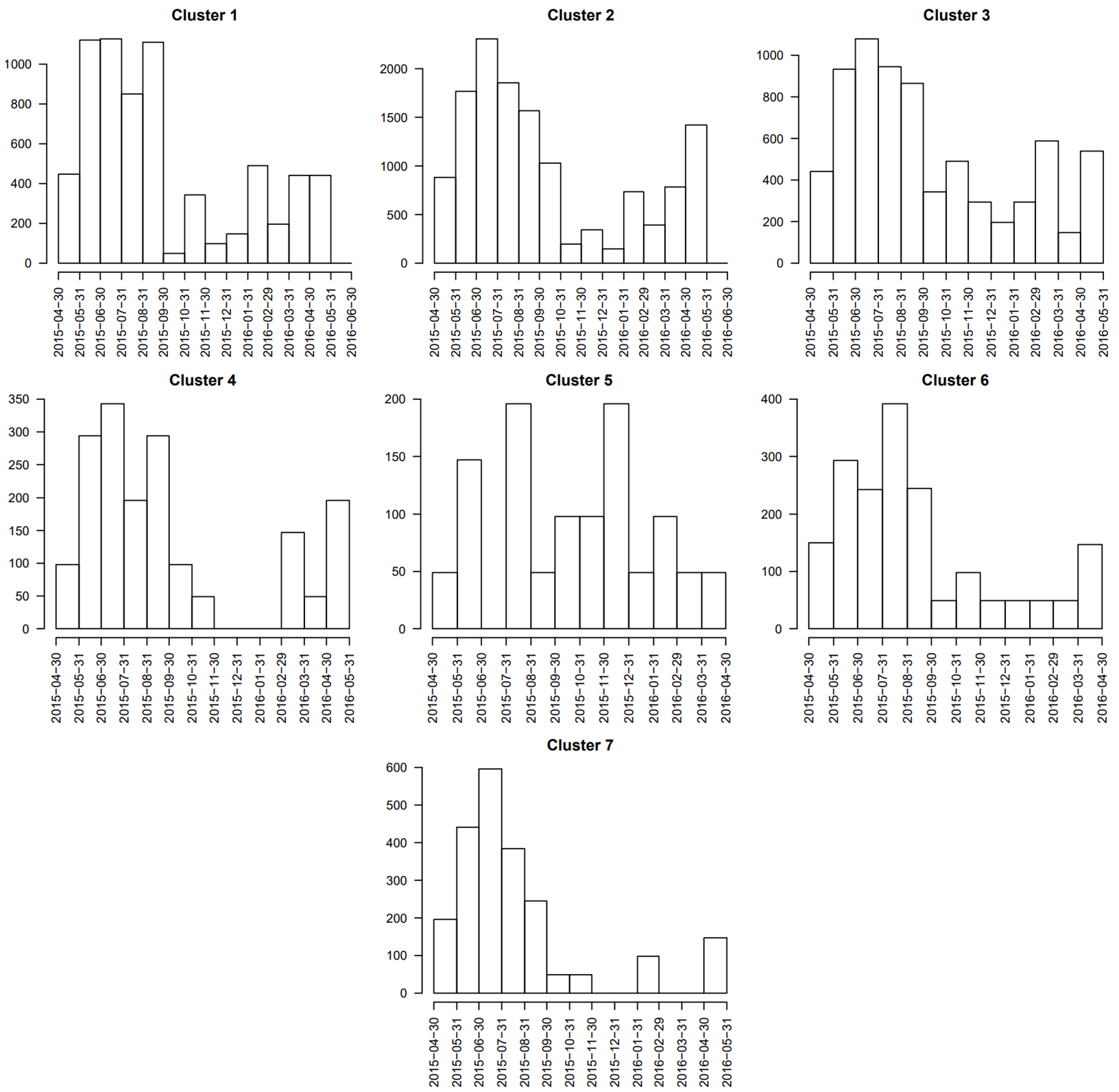}
\caption{The histogram of the dates when the vehicles of different clusters have passed through the preserve.}
\label{figure_hist_dates_carsInClusters}
\end{figure*}

In the following, we analyze the trajectories of every cluster:
\begin{itemize}
\item Clusters 1, 6, and 7 do not visit ranger base which makes sense because according to Fig. \ref{figure_heatmap_n_cars_cluster_vs_types}, they do not include any ranger.
\item Cluster 5 mostly covers all the preserve, especially observed by looking at the mean trajectory. This makes sense because according to Fig. \ref{figure_heatmap_n_cars_cluster_vs_types}, this cluster mostly includes rangers who are responsible to inspect in all regions of the preserve.
\item \textbf{Cluster 5, which mostly includes the rangers, is missing visiting the regions in the east of preserve. This is especially observed by the mean trajectory. This gives the abusers an opportunity to do suspicious activities in the east of the preserve.} This coincides with our observation in type-7 vehicles in Fig. \ref{figure_trajectories_cars_of_types}.
\item Clusters 2 and 3 almost visit all regions which makes sense because according to Fig. \ref{figure_heatmap_n_cars_cluster_vs_types}, they include vehicles from all clusters. However, the patterns of clusters 2 and 3 are different from the pattern of cluster 5 in terms of the mean trajectory. The mean trajectories of clusters 2 and 3 do not cover as much area as the mean trajectory of cluster 5. This is because they are not merely containing rangers.
\item \textbf{Except for clusters 2 and 3, all clusters do not concentrate on the east. This is strange. Also, by noticing that trajectories of cluster 5 (rangers) do not concentrate on the east, we can conclude that vehicles visiting the east in clusters 2 and 3 are not rangers. This gives us an alert that some non-ranger vehicles have visited the east a lot while the rangers have not inspected there. Hence, some suspicious activities might have happened there.}
\end{itemize}

\subsubsection{Analysis of Traffic Time Based on Traffic Clusters}

In order to analyze the data of traffic in terms of time, we plot the histogram of the dates when the vehicles of different clusters have passed through the preserve. These histograms for different clusters are shown in Fig. \ref{figure_hist_dates_carsInClusters}.
In the following, we analyze the histograms:
\begin{itemize}
\item Cluster 5 visits the preserve at various times. This makes sense because this cluster includes rangers and the rangers are supposed to visit the preserve regularly.
\item \textbf{All clusters except cluster 5 mostly have visited the preserve around June 2015. This might be destructive to the nature preserve or might have had bad effects on the quality of life of flora and fauna in the preserve.} This observation coincides with our observation from Fig. \ref{figure_hist_dates_carsInTypes}.
\end{itemize}

\begin{table*}[!t]
\setlength\extrarowheight{5pt}
\centering
\scalebox{0.75}{    
\begin{tabular}{l | l}
\hline
\hline
\textbf{Chemical} & \textbf{Characteristics}\\
\hline
\hline
Appluimonia & Airborne odor, not serious injury, influences life quality\\
\hline
Chlorodinine & Corrodes body tissues, harmful if inhaled or swallowed, used for sterilizing\\
\hline
Methylosmolene & Volatile organic solvent, toxic side effects, must be neutralized before disposal\\
\hline
AGOC-3A & Environment friendly solvents, less harmfull, low Volatile Organic Compound (VOC)\\
\hline
\hline
\end{tabular}%
}
\caption{Chemicals recorded by the sensors.}
\label{table_chemicals}
\end{table*}

\begin{table*}[!t]
\setlength\extrarowheight{5pt}
\centering
\scalebox{0.75}{    
\begin{tabular}{l | c | l}
\hline
\hline
\textbf{Company} & \textbf{Coordinate} & \textbf{Production}\\
\hline
\hline
Roadrunner Fitness Electronics & (89,27) & Fitness trackers, heart rate monitors, sport-related products\\
\hline
Kasios Office Furniture & (90,21) & Metal and composite-wood office furniture\\
\hline
Radiance ColourTek & (109,26) & Solvent-based metallic flake paints\\
\hline
Indigo Sol Boards & (120,22) & Skate-boards and snow-boards\\
\hline
\hline
\end{tabular}%
}
\caption{Manufacturing factories near the nature preserve.}
\label{table_factories}
\end{table*}

\section{Analysis of Impact of Surrounding Factories}

\subsection{Manufacturing Factories and the Sensors}

The data subset 2 provides us the data of nine sensors recording four different chemicals. The nine sensors are located at the south of the preserve. These sensors are depicted in Fig. \ref{figure_map}. The four chemicals recorded by these sensors are Appluimonia, Chlorodinine, Methylosmolene, and AGOC-3A. The characteristics of these chemicals are mentioned in Table \ref{table_chemicals}. As reported in this table, Chlorodinine and Methylosmolene are more dangerous chemicals.

Four manufacturing factories are also located in the south of the preserve, surrounded by the nine sensors. The location of these plants are also shown in Fig. \ref{figure_map}. The coordinates and the products of these factories are reported in Table \ref{table_factories}. As seen in this table, the two factories Kasios Office Furniture and Radiance ColourTek seem to be more risky to environment compared to Roadrunner Fitness Electronics  and Indigo Sol Boards.

\begin{figure*}[!t]
\centering
\includegraphics[width=5in]{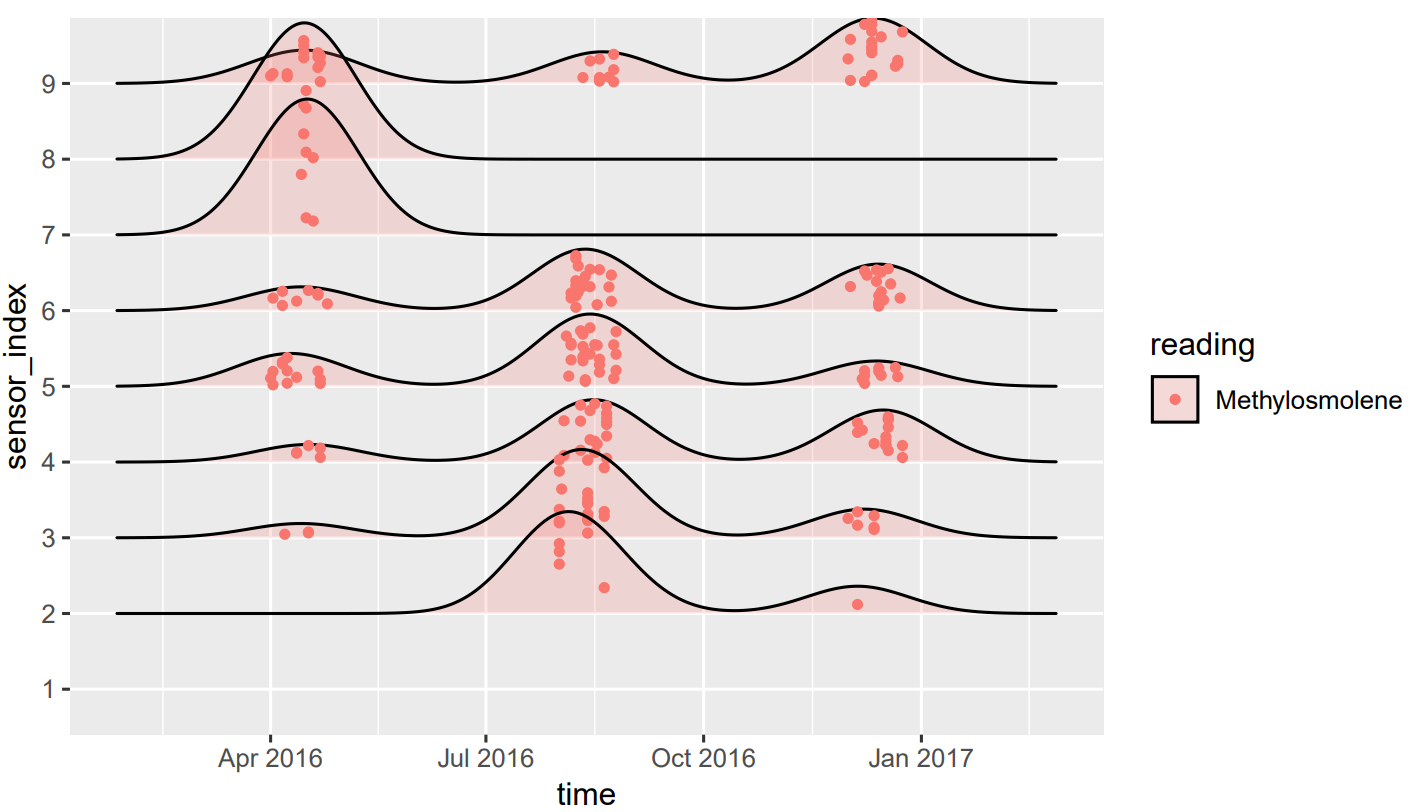}
\caption{The ridge plot (Joyplot) of the sensor failures in recording from the chemicals.}
\label{figure_sensor_performance_ridgePlot}
\end{figure*}

\subsection{Analysis of Performance of Sensors}

By carefully observing the data of sensors, we figure out that the sensors have some lacks and failures in their recordings. In other words, at every time slot, the sensor is supposed to record the four chemical values. For every sensor, we check whether all the chemicals have been recorded or not. The ridge plot (Joyplot) of the sensor failures in recording from the chemicals is shown in Fig. \ref{figure_sensor_performance_ridgePlot}. As can be seen in this plot:
\begin{itemize}
\item \textbf{Mostly Methylosmolene is not recorded at some time slots.}
\item \textbf{Sensor 1 does not have any failure but other sensors have failed in recording Methylosmolene at some times.}
\item The more strange fact is that \textbf{these failures have occurred mostly at three time periods, i.e., April and June 2016, August and September 2016, and December 2016. The sensors have failed at the same times. Probably this has happened because of a problem in the electricity or supplies of the sensors.}
\end{itemize}
Note that some more analysis of the performance of sensors is done in the following sections.

\begin{figure*}[!t]
\centering
\includegraphics[width=6.5in]{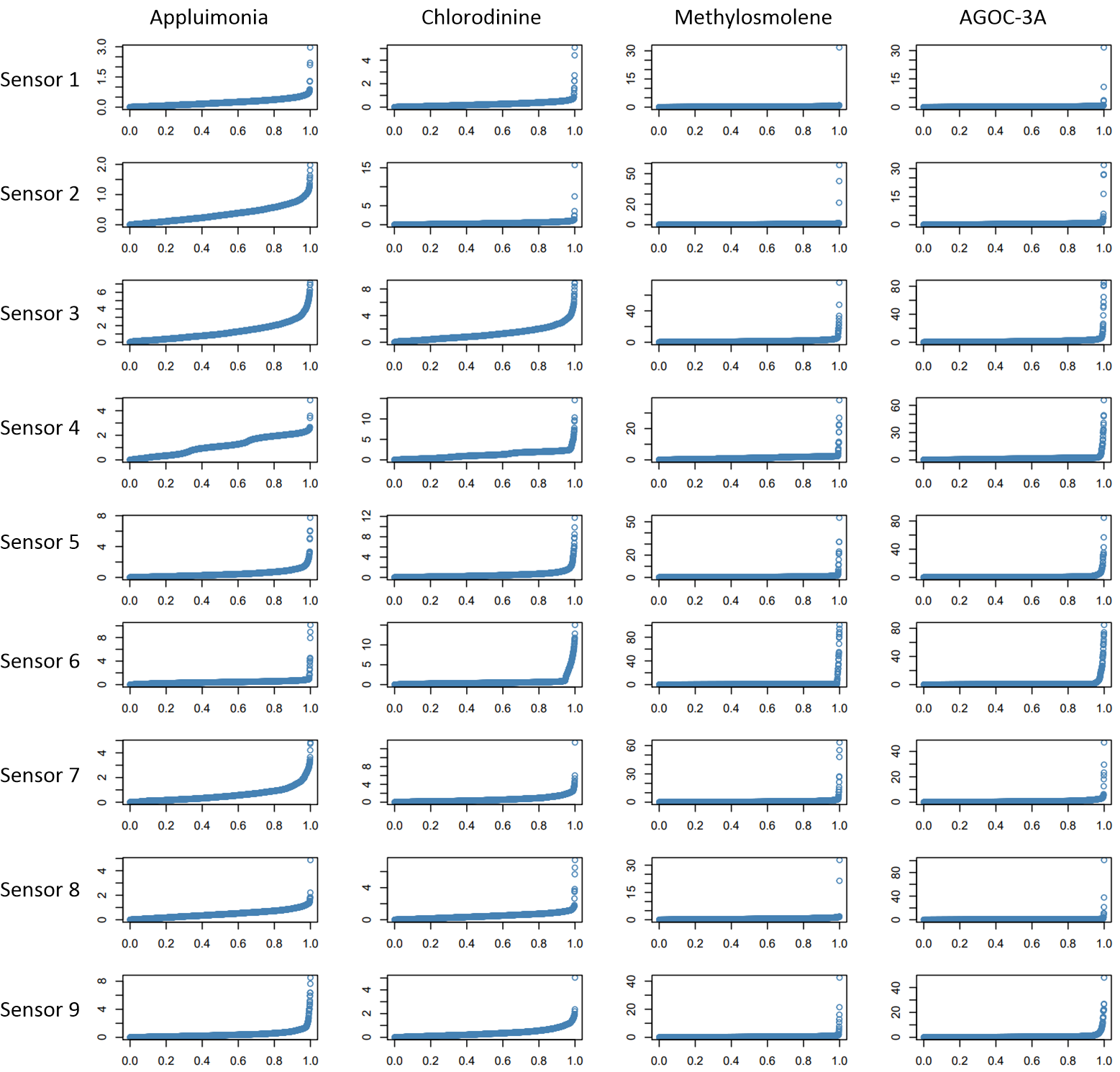}
\caption{The quantile plots of the sensor records from the chemicals. Rows and columns correspond to the sensors and chemicals, respectively, with the order mentioned in Fig. \ref{figure_map} and Table \ref{table_chemicals}.}
\label{figure_SensorChemicalGrid_quantile}
\end{figure*}

\subsection{Analysis of Chemicals}

In this section, we analyze the recorded chemicals by the nine sensors. The metric of records is parts per million.

\subsubsection{Distribution of The Recorded Chemicals}

The quantile plot of the recorded of every chemical by every sensor is depicted in Fig. \ref{figure_SensorChemicalGrid_quantile}.
In the following, we analyze the quantile plots:
\begin{itemize}
\item For some chemicals and sensors, most of recordings are very small as it is expected according to the rules of environmental protection. However, we can see some outliers which are showing high values of the chemicals. For example, \textbf{for all sensors, we have very high values (30 to 80 parts per million) of Methylosmolene and AGOC-3A.}
\item We also have some outliers (relatively high values) of Appluimonia and Chlorodinine recorded by different sensors. For example, \textbf{sensor 2 has recorded some high values of Chlorodinine.}
\item For most of the sensors, the recordings of Appluimonia and Chlorodinine are \textbf{very skewed around zero}. This makes sense because the values are supposed to be small. 
\item There is some \textbf{strange granularity in distribution of Appluimonia and Chlorodinine recorded by sensor 4. This probably shows some sensor failures in the recording or some strange patterns of chemicals in that region.}
\item \textbf{In the recorded data by sensor 6, we can see some strange breaking knee points. This again might be because of failures of the sensor or the odd patterns of chemical values. However, as this has happened for all chemicals, the hypothesis of sensor problem is stronger.}
\end{itemize}

\subsubsection{The Recorded Chemicals Over Time}

\begin{figure*}[!t]
\centering
\includegraphics[width=6.5in]{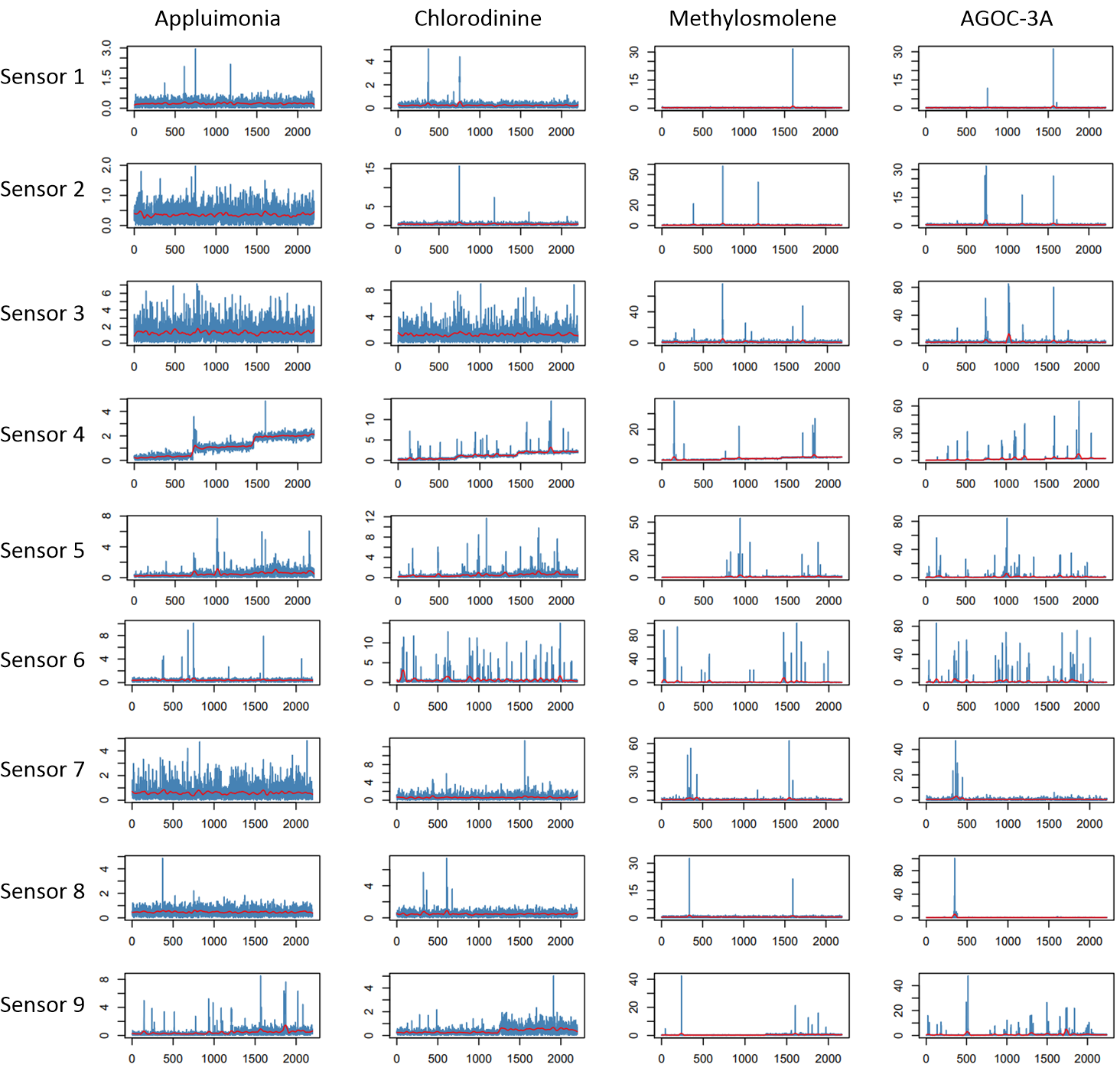}
\caption{The time series plots of the sensor records from the chemicals. Rows and columns correspond to the sensors and chemicals, respectively, with the order mentioned in Fig. \ref{figure_map} and Table \ref{table_chemicals}. The red curves are the smoothed curves of the time series with time span $5\%$.}
\label{figure_SensorChemicalGrid_timeSeries}
\end{figure*}

The time interval of the recordings of the sensors is from April 1st, 2016 to December 31st, 2016. The recordings of the four chemicals by the nine sensors are shown as time series in Fig. \ref{figure_SensorChemicalGrid_timeSeries}. This figure also shows the smoothed curves of the time series by red color.
We analyze the different time series in the following:
\begin{itemize}
\item Again, we see that \textbf{for all sensors, we have very high values (30 to 80 parts per million) of Methylosmolene and AGOC-3A at some times. These picks have happened at some specific times and not all times.} The more accurate report of these times are presented in the next sections.
\item We can also see some \textbf{high values of Appluimonia and Chlorodinine at some specific times.} The more accurate report of these times are presented in the next sections.
\item There is some \textbf{strange granularity in the time series of Appluimonia and Chlorodinine recorded by sensor 4. But this issue does not exist for records of Methylosmolene and AGOC-3A by this sensor, suggesting that this might not be because of sensor failure. According to the increasing behaviour of records of Appluimonia and Chlorodinine by sensor 4, we can conclude that these chemicals have been accumulated around the regions covered by sensor 4.}
\item \textbf{The records of sensor 6 for all four chemicals are more frequently having high values, giving the clue that some odd chemical accumulation might have happened at different months of 2016. This might explain the reason of breaking knee points in the quantile plots of this sensor in Fig. \ref{figure_SensorChemicalGrid_quantile}.}
\end{itemize}

\subsubsection{Unusually High Values of Chemicals}

\begin{figure*}[!t]
\centering
\includegraphics[width=5.5in]{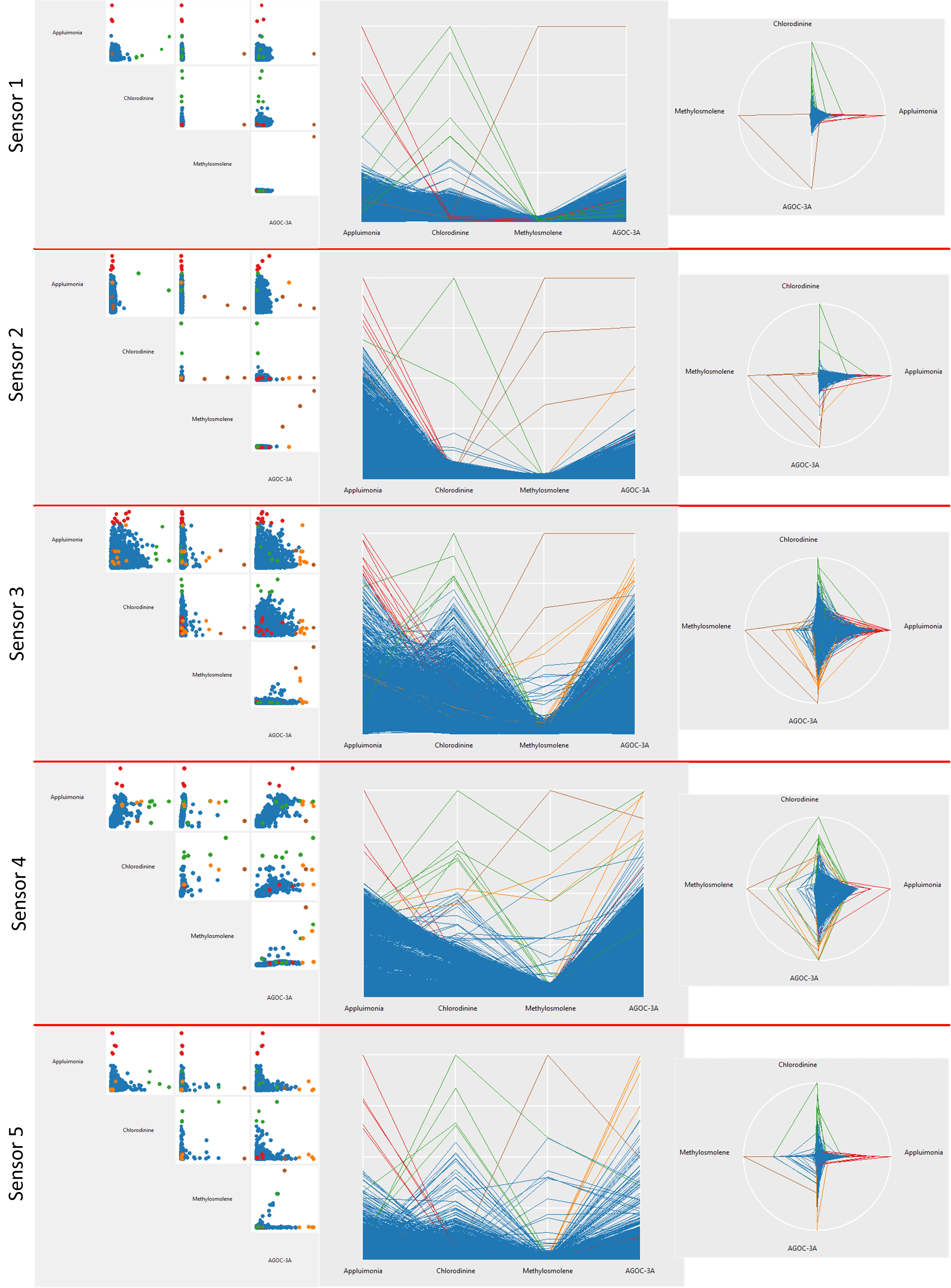}
\caption{The pairs plot, parallel axes plot, and star plot for chemical recordings by sensors (continue in Fig. \ref{figure_loon_sensors_2}). In the parallel axes and star plots, the data is scaled by variable. The red, green, brown, and orange points indicate the large values of Appluimonia, Chlorodinine, Methylosmolene, and AGOC-3A, respectively. Note that in some cases, a combination of colors show the large values for a  chemical.}
\label{figure_loon_sensors_1}
\end{figure*}

\begin{figure*}[!t]
\centering
\includegraphics[width=5.5in]{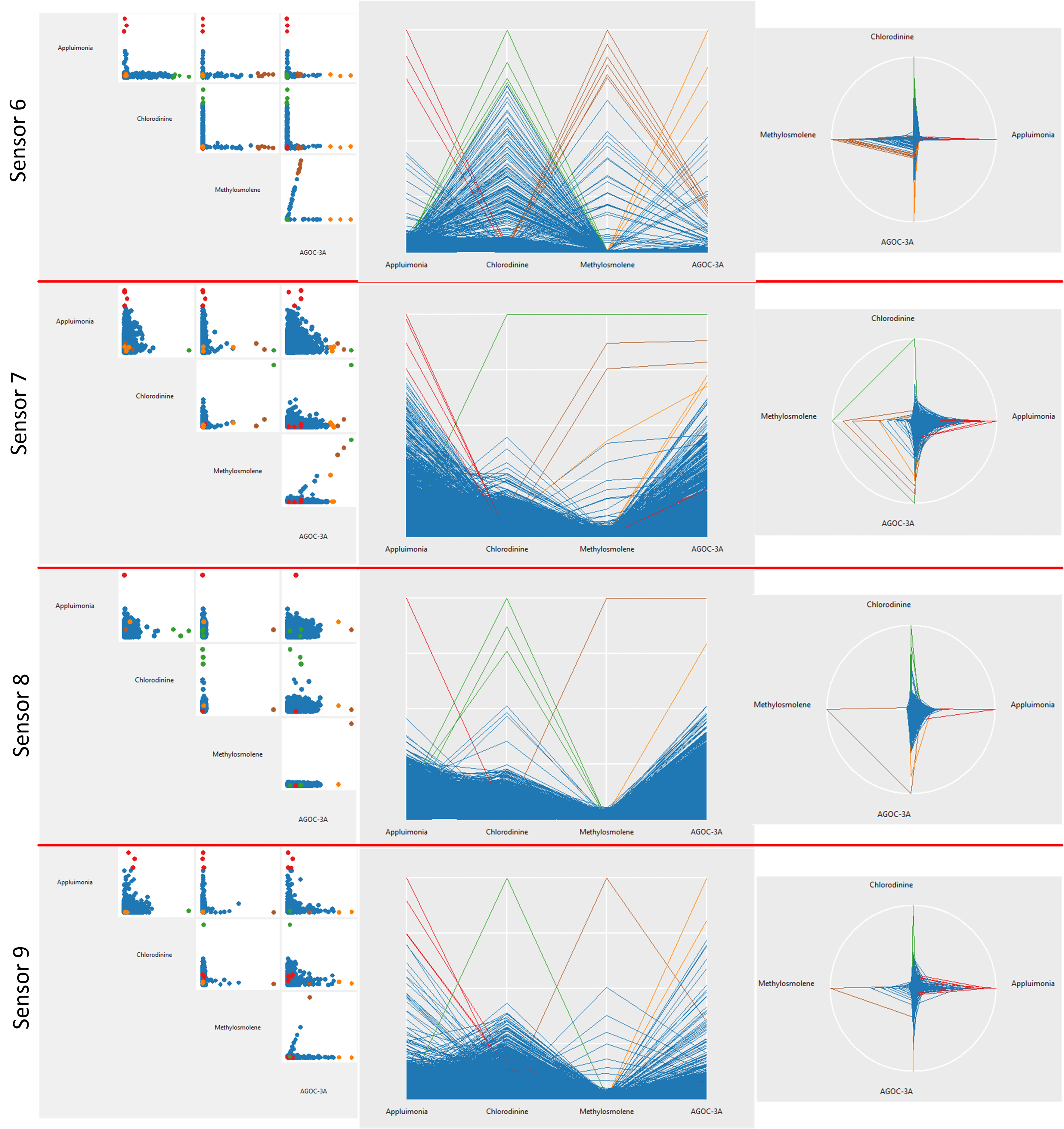}
\caption{Continue of Fig. \ref{figure_loon_sensors_1}.}
\label{figure_loon_sensors_2}
\end{figure*}

Using the interactive ``loon'' package in the R programming language \cite{web_loon}, we plot the pairs plot (scatter plot matrix), parallel axes plot, and star plot of the recorded chemicals while the plots are linked together. These plots are shown in figures \ref{figure_loon_sensors_1} and \ref{figure_loon_sensors_2}. For every sensor, the three plots are shown.

In the plots of every sensor, we highlight the unusually high values of the four chemicals. We use different colors for the different chemicals. The utilized colors are red, green, brown, and orange for large values of Appluimonia, Chlorodinine, Methylosmolene, and AGOC-3A, respectively. 

Afterwards, using the manually determined colors, we label the recorded data as usual and high values for each chemical recorded by every sensor. Therefore, we have the indices of high chemical values for every chemical in every sensor record. These indices are used in the following sections where we analyze the high values of chemicals.

The \textbf{correlation of the recorded chemical values} can be discussed here. Here, the chemicals are the dimensions of data. The correlation of the variables (dimensions) can be analyzed using either scatter plots, parallel axes, or the star plot. In scatter plot, if the points almost form a line with positive/negative slope, the correlation is positive/negative. In the parallel axes or the star plot, if the high/low values of a variable correspond to the high/low values of the other variable, those two variables are positively correlated; however, if the high/low values of a variable correspond to the low/high values of another one, we have negative correlation. 

From the plots, we can see that:
\begin{itemize}
\item In \textbf{sensor 1}, \textit{if we ignore some rare outliers (high values of chemicals)}, we can say that \textbf{Appluimonia and Methylosmolene, Chlorodinine and Methylosmolene, and Methylosmolene and AGOC-3A are almost independent}. That is because given a variable, the value of other variable does not change in the scatter plot.
\item In \textbf{sensor 2}, \textit{if we ignore some rare outliers (high values of chemicals)}, we can say that \textbf{Appluimonia and Chlorodinine, Appluimonia and Methylosmolene, Chlorodinine and Methylosmolene are almost independent}. That is because given a variable, the value of other variable does not change in the scatter plot.
\item In \textbf{sensor 3}, the \textit{correlation of Appluimonia and Chlorodinine, the correlation of Appluimonia and AGOC-3A, and the correlation of Chlorodinine and AGOC-3A are relatively small} compared to other correlations because the points of the scatter plot roughly (but not completely) cover many parts of the plot, similar to the scatter plot of two independent variables.
\item In \textbf{sensor 4}, the variables \textbf{Appluimonia and Chlorodinine, Appluimonia and AGOC-3A, and Chlorodinine and AGOC-3A are positively correlated}. 
\item In \textbf{sensor 5}, the variables \textbf{Appluimonia and Chlorodinine, Appluimonia and AGOC-3A, and Chlorodinine and AGOC-3A are negatively correlated}. Also, we can see from the pattern of scatter plots that \textbf{Appluimonia and Methylosmolene, Chlorodinine and Methylosmolene, and Methylosmolene and AGOC-3A have some interesting and specific dependence.}
\item In \textbf{sensor 6}, we can see from the pattern of scatter plots that \textbf{all the chemicals have some interesting and specific dependence.}
\item In \textbf{sensor 7}, \textit{if we ignore some rare outliers (high values of chemicals)}, we can say that \textbf{Appluimonia and Chlorodinine, and Chlorodinine and AGOC-3A are almost independent}. Also, we can see that \textbf{Appluimonia and AGOC-3A are almost negatively correlated.}
\item In \textbf{sensor 8}, \textit{if we ignore some rare outliers (high values of chemicals)}, we can say that \textbf{Appluimonia and Methylosmolene, Chlorodinine and Methylosmolene, and Methylosmolene and AGOC-3A are almost independent}.
\item In \textbf{sensor 9}, we can say that \textbf{Appluimonia and AGOC-3A are almost negatively correlated}. Also, we can see from the pattern of scatter plots that \textbf{Methylosmolene and AGOC-3A have some interesting and specific dependence.}
\end{itemize}

\subsection{Analysis of Meteorological Data and Factories}

\subsubsection{Distribution of The Meteorological Data}

The dataset provides us the meteorological data for a time period. The given meteorological data include both wind direction and speed. The wind speed is in meters per second and the wind direction is where the wind is originating from, using a north-referenced azimuth bearing \cite{rutstrum2000wilderness} where 360/000 is true north. To better explain the wind direction in this standard, note that: 
\begin{itemize}
\item Wind direction 0 (or 360) means wind is blowing from north to south.
\item Wind direction 90 means wind is blowing from east to west.
\item Wind direction 180 means wind is blowing from south to north.
\item Wind direction 270 means wind is blowing from west to east.
\end{itemize}

\begin{figure}[!t]
\centering
\includegraphics[width=2.5in]{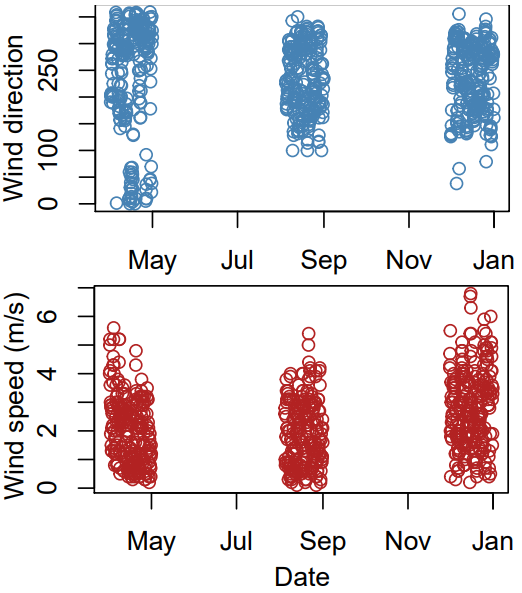}
\caption{The meteorological data provided for three different months, i.e., April 2016, August 2016, and December 2016.}
\label{figure_meteorological_allDates}
\end{figure}

We plot the scatter plot of the meteorological data versus time in Fig. \ref{figure_meteorological_allDates}. From this figure, we find out that the given meteorological data contain information for three months, i.e., April 2016, August 2016, and December 2016.

\begin{figure*}[!t]
\centering
\includegraphics[width=6in]{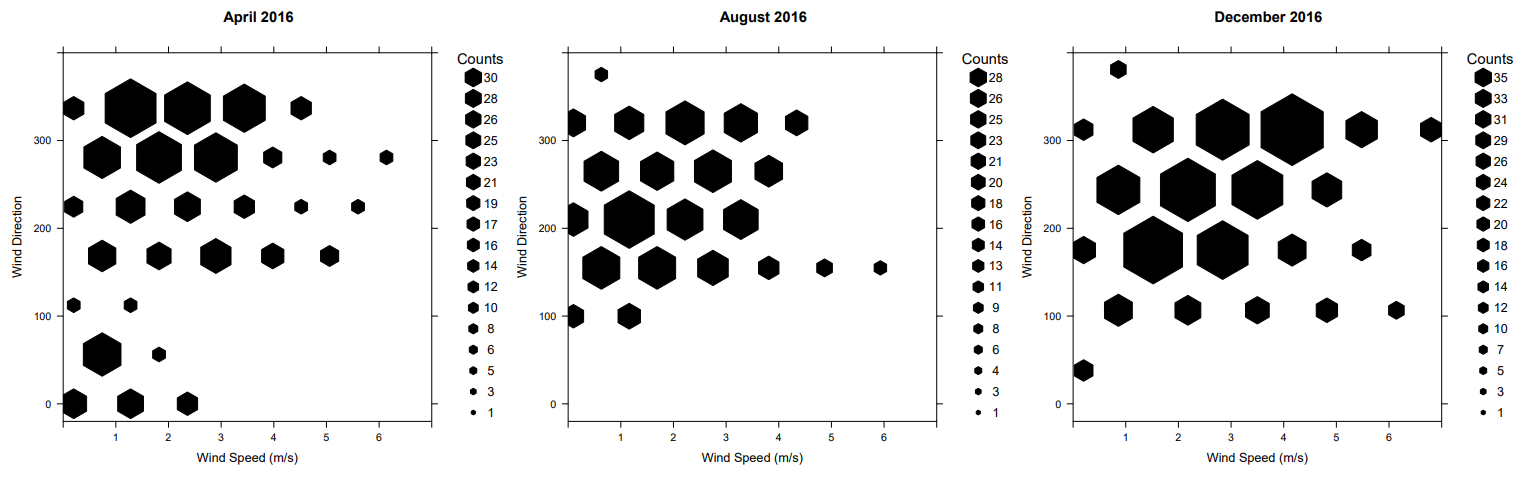}
\caption{The hexagonal 2D histograms of meteorological data, i.e., the wind direction versus the wind speed during months April, August, and December 2016.}
\label{figure_meteorological_hexbinplot}
\end{figure*}

As the meteorological data contain April 2016, August 2016, and December 2016, we separate the meteorological data of these three months from the dataset. Figure \ref{figure_meteorological_hexbinplot} shows the hexagonal 2D histograms of meteorological data, i.e., the wind direction versus the wind speed during the months April, August, and December 2016. From these histograms, we can see the density and concentration of the wind direction and speed in the three months:
\begin{itemize}
\item In \textbf{April 2018}, the wind mostly blows \textbf{from north to south} and the speed is mostly \textbf{1 m/s}. Moreover, in some cases, we have wind with speed \textbf{less than 1 to 2 m/s} blowing \textbf{from east to west}.
\item In \textbf{August 2018}, the wind mostly blows \textbf{from south-west to north-east} and the speed is mostly \textbf{1 to 2 m/s}.
\item In \textbf{December 2018}, the wind mostly blows either \textbf{from south to north} with speed \textbf{1.5 m/s} or \textbf{from north-west to south-east} with speed \textbf{4 m/s}.
\end{itemize}

\subsubsection{The Meteorological Data During Time}

\begin{figure*}[!t]
\centering
\includegraphics[width=6in]{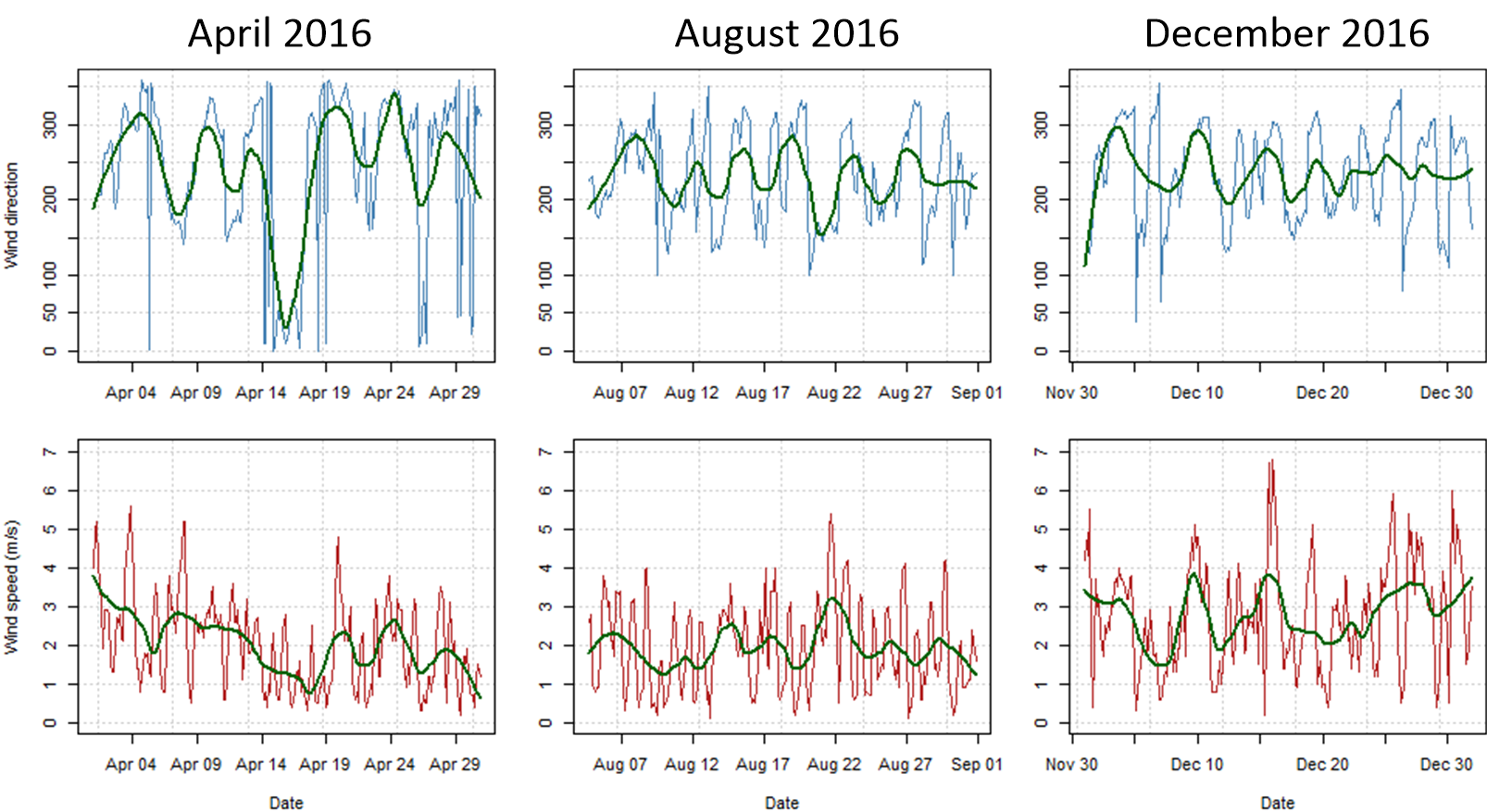}
\caption{The time series of meteorological data, i.e., wind direction and speed during months April 2016 (first column), August 2016 (second column), and December 2016 (third column). The green curves are the smoothed curves of the time series with span $20\%$.}
\label{figure_meteorological_time_series}
\end{figure*}

It is also useful to show the meteorological data in terms of time because Fig. \ref{figure_meteorological_hexbinplot} does not encode the detailed time within the months. Figure \ref{figure_meteorological_time_series} shows the times series of meteorological data, including both wind direction and speed, for the three months. We can conclude from this figure that:
\begin{itemize}
\item For \textbf{April 2016}, on average, the wind direction is \textbf{from north to south} but \textbf{in the middle of April, we have many different directions of wind}. The speed is around \textbf{1 or 2 m/s} but is \textbf{slowly decreasing}.
\item For \textbf{August 2016}, on average, the wind blows \textbf{from south-west to north-east} with speed \textbf{1 or 2 m/s}.
\item For \textbf{December 2016}, on average, the wind blows \textbf{from south-west to north-east} with speed \textbf{2 or 3 m/s}. The wind direction and speed have \textbf{fluctuation} in this month.
\end{itemize}

We summarize the meteorological information extracted from the Figures \ref{figure_meteorological_hexbinplot} and \ref{figure_meteorological_time_series} in Table \ref{table_meteorological}.

\begin{table*}[!t]
\setlength\extrarowheight{5pt}
\centering
\scalebox{0.7}{    
\begin{tabular}{l | l | l}
\hline
\hline
\textbf{Month} & \textbf{During the month} & \textbf{On average}\\
\hline
\hline
April & north to south, 1 to 2 m/s $+$ some fluctuation in mid April & north to south, 1 or 2 m/s\\
\hline
August & south-west to north-east, 1 to 2 m/s & south-west to north-east, 1 to 2 m/s\\
\hline
December & south to north, 1.5 m/s $+$ north-west to south-east, 4 m/s & south-west to north-east, 2 or 3 m/s\\
\hline
\hline
\end{tabular}%
}
\caption{Meteorological information in April, August, and December 2016.}
\label{table_meteorological}
\end{table*}

\subsubsection{Analysis of High Chemical Values with the Meteorological Data}

\begin{table}[!t]
\setlength\extrarowheight{5pt}
\centering
\scalebox{0.6}{    
\begin{tabular}{c | l | c | l}
\hline
\hline
\textbf{Sensor} & \textbf{Time} & \textbf{Wind Direction} & \textbf{Suspect factories} \\
\hline
\hline
1 & April 26 & $\downarrow$ & --\\
1 & August 1  & $\nearrow$ & --\\
1 & August 20 & $\leftarrow$ & all\\
\hline
2 & April 4 & $\rightarrow$ & --\\
2 & April 14 & $\downarrow$ & --\\
2 & April 29 & $\leftarrow$ & all\\
2 & August 1 & $\nearrow$ & --\\
2 & December 7 & $\leftarrow$ & all\\
\hline
3 & April 6 & $\uparrow$ & Roadrunner, Kasios\\
3 & April 20 & $\downarrow$ & --\\
3 & April 26 & $\downarrow$ & --\\
3 & August 1 & $\uparrow$ & Roadrunner, Kasios\\
3 & August 20 & $\nwarrow$ & Roadrunner, Kasios\\
3 & August 25 & $\uparrow$ & Roadrunner, Kasios\\
3 & December 18 & $\uparrow$ & Roadrunner, Kasios\\
\hline
4 & August 1 & $\nearrow$ & --\\
4 & December 8 & $\nwarrow$ & all\\
\hline
5 & August 13 & $\downarrow$ & --\\
5 & December 8 & $\nearrow$ & Roadrunner, Kasios\\
5 & December 28 & $\nearrow$ & Roadrunner, Kasios\\
\hline
6 & April 29 & $\downarrow$ & Roadrunner, Radiance\\
6 & August 1 & $\uparrow$ & --\\
6 & December 8 & $\nearrow$ & Kasios\\
\hline
7 & April 29 & $\downarrow$ & Roadrunner, Kasios\\
7 & August 1 & $\uparrow$ & --\\
7 & December 22 & $\nearrow$ & --\\
7 & December 29 & $\nwarrow$ & --\\
\hline
8 & April 15 & $\downarrow$ & Roadrunner, Kasios\\
\hline
9 & December 5 & $\nearrow$ & all\\
9 & December 18 & $\uparrow$ & Radiance, Indigo\\
9 & December 25 & $\uparrow$ & Radiance, Indigo\\
\hline
\hline
\end{tabular}%
}
\caption{Analyzing which factories are responsible for high values of Appluimonia chemical.}
\label{table_results_prediction_Appluimonia}
\end{table}

\begin{table}[!t]
\setlength\extrarowheight{5pt}
\centering
\scalebox{0.6}{    
\begin{tabular}{c | l | c | l}
\hline
\hline
\textbf{Sensor} & \textbf{Time} & \textbf{Wind Direction} & \textbf{Suspect factories} \\
\hline
\hline
1 & April 16 & $\downarrow$ & --\\
1 & August 1  & $\nearrow$ & --\\
\hline
2 & August 1 & $\nearrow$ & --\\
2 & August 20 & $\leftarrow$ & all\\
\hline
3 & August 1 & $\nearrow$ & --\\
3 & December 1 & $\nwarrow$ & all\\
3 & December 5 & $\downarrow$ & --\\
3 & December 10 & $\rightarrow$ & --\\
3 & December 29 & $\nwarrow$ & all\\
\hline
4 & April 5 & $\uparrow$ & Roadrunner, Kasios\\
4 & December 5 & $\nwarrow$ & Radiance, Indigo\\
4 & December 18 & $\uparrow$ & Roadrunner, Kasios\\
4 & December 25 & $\uparrow$ & Roadrunner, Kasios\\
\hline
5 & August 15 & $\uparrow$ & all\\
5 & December 12 & $\nearrow$ & Roadrunner, Kasios\\
5 & December 22 & $\nearrow$ & Roadrunner, Kasios\\
\hline
6 & April 4 & $\rightarrow$ & Roadrunner, Kasios\\
6 & April 10 & $\rightarrow$ & Roadrunner, Kasios\\
6 & April 26 & $\rightarrow$ & Roadrunner, Kasios\\
6 & December 22 & $\rightarrow$ & Roadrunner, Kasios\\
\hline
7 & December 5 & $\nwarrow$ & --\\
\hline
8 & April 14 & $\downarrow$ & Roadrunner, Kasios\\
8 & April 26 & $\downarrow$ & Roadrunner, Kasios\\
\hline
9 & December 18 & $\rightarrow$ & all\\
\hline
\hline
\end{tabular}%
}
\caption{Analyzing which factories are responsible for high values of  Chlorodinine chemical.}
\label{table_results_prediction_Chlorodinine}
\end{table}

\begin{table}[!t]
\setlength\extrarowheight{5pt}
\centering
\scalebox{0.6}{    
\begin{tabular}{c | l | c | l}
\hline
\hline
\textbf{Sensor} & \textbf{Time} & \textbf{Wind Direction} & \textbf{Suspect factories} \\
\hline
\hline
1 & December 8 & $\nwarrow$ & --\\
\hline
2 & April 16  & $\nwarrow$ & Roadrunner, Kasios\\
2 & August 1  & $\nearrow$ & --\\
2 & August 20  & $\leftarrow$ & all\\
\hline
3 & August 1  & $\nearrow$ & --\\
3 & December 12  & $\nwarrow$ & Roadrunner, Kasios\\
\hline
4 & April 6  & $\uparrow$ & Roadrunner, Kasios\\
\hline
5 & August 10  & $\nwarrow$ & Radiance, Indigo\\
\hline
6 & April 3  & $\rightarrow$ & Roadrunner, Kasios\\
6 & April 10  & $\rightarrow$ & Roadrunner, Kasios\\
6 & December 3  & $\rightarrow$ & Roadrunner, Kasios\\
6 & December 9  & $\nearrow$ & Kasios\\
\hline
7 & April 15  & $\downarrow$ & Roadrunner, Kasios\\
7 & December 5  & $\uparrow$ & --\\
\hline
8 & April 15  & $\downarrow$ & Roadrunner, Kasios\\
\hline
9 & April 11  & $\uparrow$ & Radiance, Indigo\\
\hline
\hline
\end{tabular}%
}
\caption{Analyzing which factories are responsible for high values of  Methylosmolene chemical.}
\label{table_results_prediction_Methylosmolene}
\end{table}

\begin{table}[!t]
\setlength\extrarowheight{5pt}
\centering
\scalebox{0.6}{    
\begin{tabular}{c | l | c | l}
\hline
\hline
\textbf{Sensor} & \textbf{Time} & \textbf{Wind Direction} & \textbf{Suspect factories} \\
\hline
\hline
1 & December 8 & $\nwarrow$ & --\\
\hline
2 & April 16 & $\nwarrow$ & all\\
2 & August 1 & $\nearrow$ & --\\
2 & August 20 & $\leftarrow$ & all\\
2 & December 8 & $\nwarrow$ & all\\
\hline
3 & August 13 & $\downarrow$ & --\\
3 & August 17 & $\uparrow$ & Roadrunner, Kasios\\
3 & August 29 & $\leftarrow$ & --\\
3 & December 12 & $\nwarrow$ & all\\
3 & December 26 & $\downarrow$ & --\\
3 & December 28 & $\uparrow$ & Roadrunner, Kasios\\
\hline
4 & April 6 & $\uparrow$ & Roadrunner, Kasios\\
4 & December 1 & $\nwarrow$ & Radiance, Indigo\\
4 & December 12 & $\nearrow$ & Roadrunner, Kasios\\
4 & December 18 & $\uparrow$ & Roadrunner, Kasios\\
\hline
5 & April 6 & $\uparrow$ & all\\
5 & August 1 & $\nearrow$ & Roadrunner, Kasios\\
5 & December 17 & $\uparrow$ & all\\
\hline
6 & April 15 & $\downarrow$ & Roadrunner, Radiance\\
\hline
7 & April 15 & $\downarrow$ & Roadrunner, Kasios\\
7 & April 19 & $\downarrow$ & Roadrunner, Kasios\\
7 & August 1 & $\nearrow$ & --\\
7 & December 15 & $\nearrow$ & --\\
\hline
8 & April 15 & $\downarrow$ & Roadrunner, Kasios\\
\hline
9 & August 12 & $\nearrow$ & all\\
9 & August 25 & $\uparrow$ & Radiance, Indigo\\
\hline
\hline
\end{tabular}%
}
\caption{Analyzing which factories are responsible for high values of  AGOC-3A chemical.}
\label{table_results_prediction_AGOC3A}
\end{table}

\begin{figure*}[!t]
\centering
\includegraphics[width=5.5in]{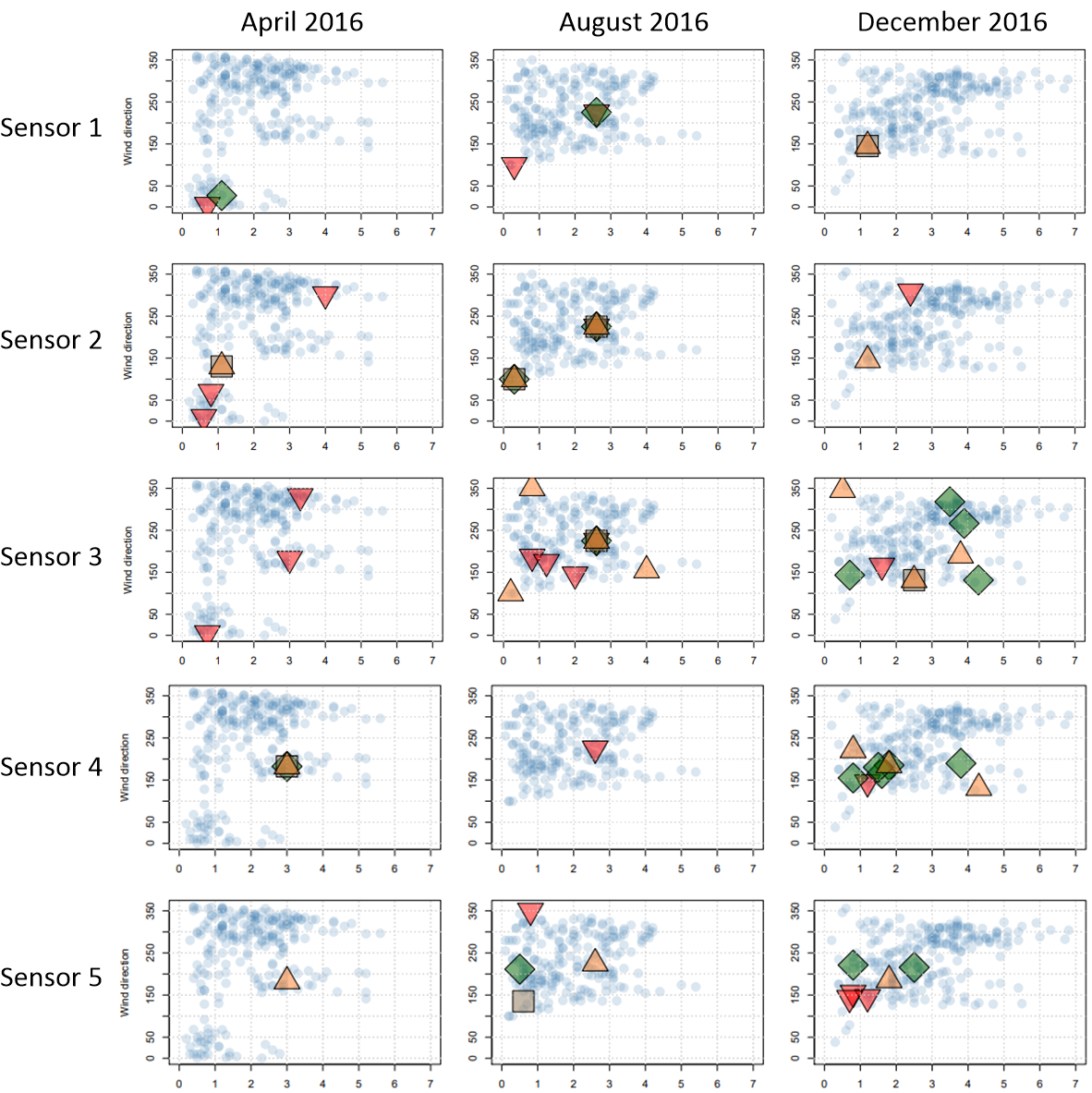}
\caption{The scatter plot of wind speed and direction while the unusual high values of chemicals recorded by the sensors are highlighted (continue in Fig. \ref{figure_meteorological_matchedDates_2}). The color of points match the used colors in Figures \ref{figure_loon_sensors_1} and \ref{figure_loon_sensors_2}. The opposite red rectangle, the green diamond, the brown rectangle, and the orange triangle correspond to high values of Appluimonia, Chlorodinine, Methylosmolene, and AGOC-3A, respectively. The small blue points correspond to the normal chemical amounts.}
\label{figure_meteorological_matchedDates_1}
\end{figure*}

\begin{figure*}[!t]
\centering
\includegraphics[width=5.2in]{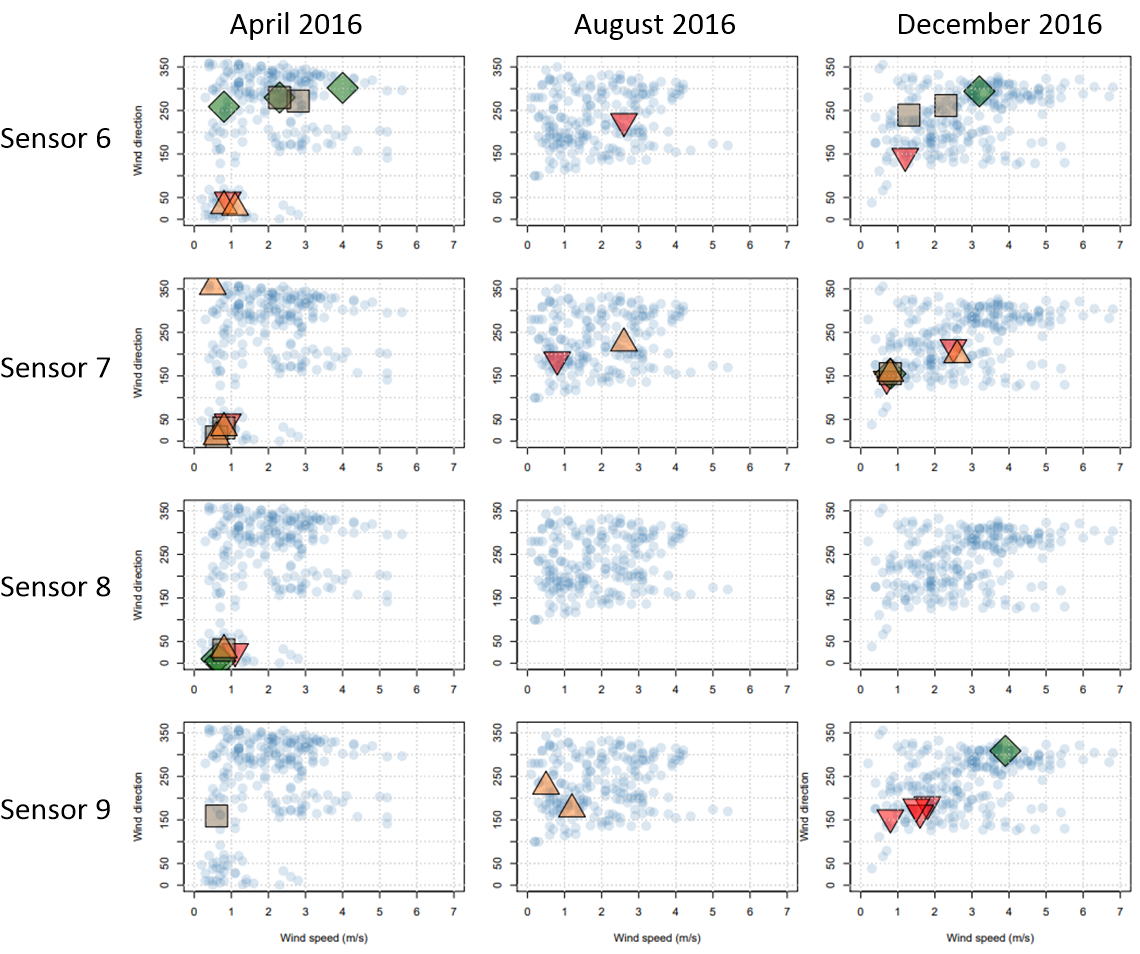}
\caption{Continue of Fig. \ref{figure_meteorological_matchedDates_1}.}
\label{figure_meteorological_matchedDates_2}
\end{figure*}

In this section, we combine the information extracted from the records of chemical values and the meteorological data in order to figure out which manufacturing factories were responsible for high values of which chemicals and when those happened. We plot the scatter plots of the meteorological data in the three months of April, August, and December in Figures \ref{figure_meteorological_matchedDates_1} and \ref{figure_meteorological_matchedDates_2}. In these figures, we do this plotting for all the nine sensors where the points in the scatter plots are color-coded by whether they are small or unusually high values of chemicals. Hence, for every sensor, we have three scatter plots for the three months where the unusual values of chemicals are shown. The colors of the unusual values of chemicals match the colors used in Figures \ref{figure_loon_sensors_1} and \ref{figure_loon_sensors_2}.

The fine values of chemicals are colored transparent (alpha-blended) blue in Figures \ref{figure_meteorological_matchedDates_1} and \ref{figure_meteorological_matchedDates_2}. Therefore, the concentration and density of the wind direction and speed can also be observed in these plots. The analysis is the same as the analysis of the concentration of data in Fig. \ref{figure_meteorological_hexbinplot}.

\begin{figure*}[!t]
\centering
\includegraphics[width=5.5in]{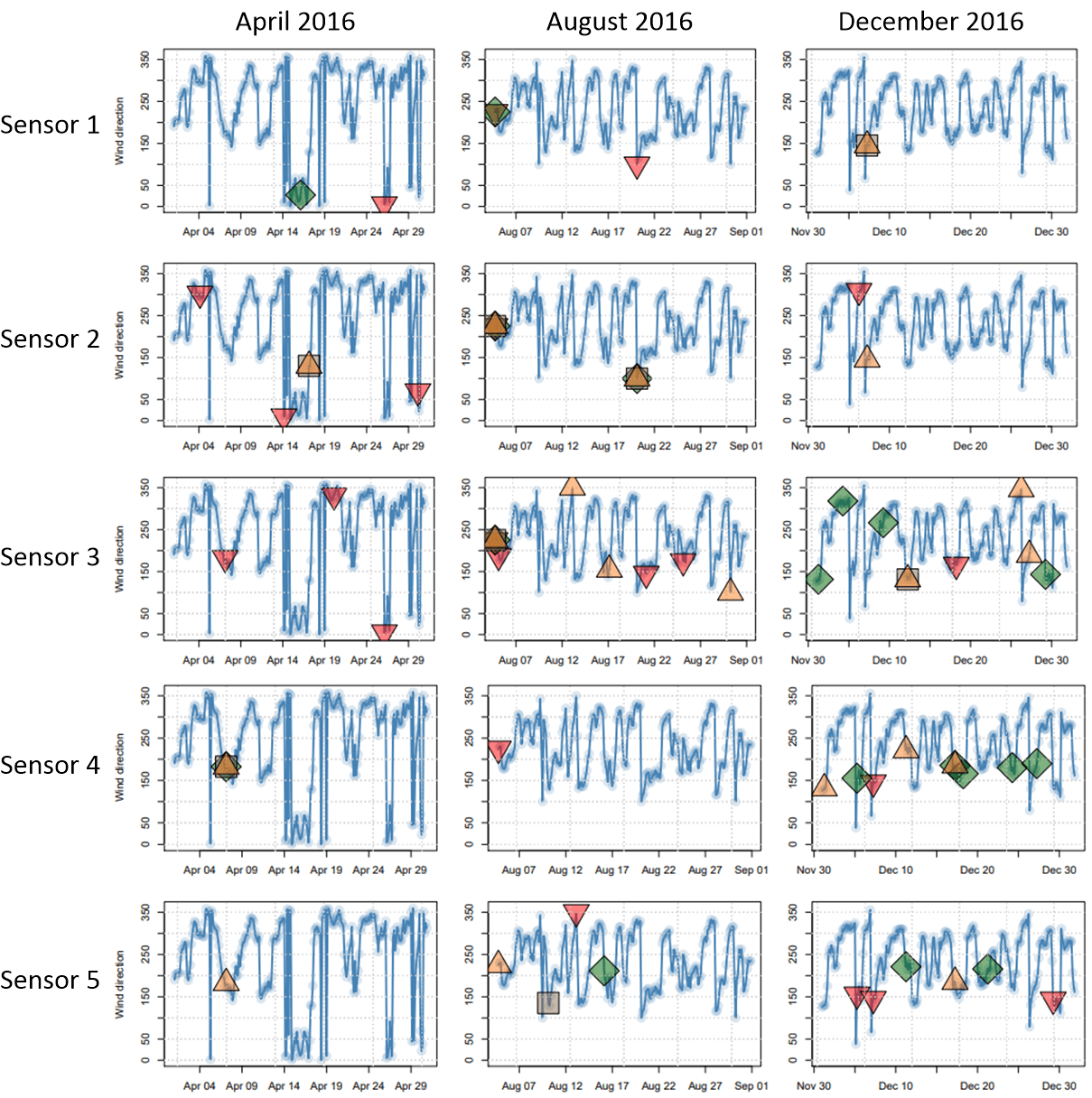}
\caption{The time series of wind direction while the unusual high values of chemicals recorded by the sensors are highlighted (continue in Fig. \ref{figure_meteorological_matchedDates_timeSeries_1}). The color of points match the used colors in figures \ref{figure_loon_sensors_1} and \ref{figure_loon_sensors_2}. The opposite red rectangle, the green diamond, the brown rectangle, and the orange triangle correspond to high values of Appluimonia, Chlorodinine, Methylosmolene, and AGOC-3A, respectively. The small blue points correspond to the normal chemical amounts.}
\label{figure_meteorological_matchedDates_timeSeries_1}
\end{figure*}

\begin{figure*}[!t]
\centering
\includegraphics[width=5.2in]{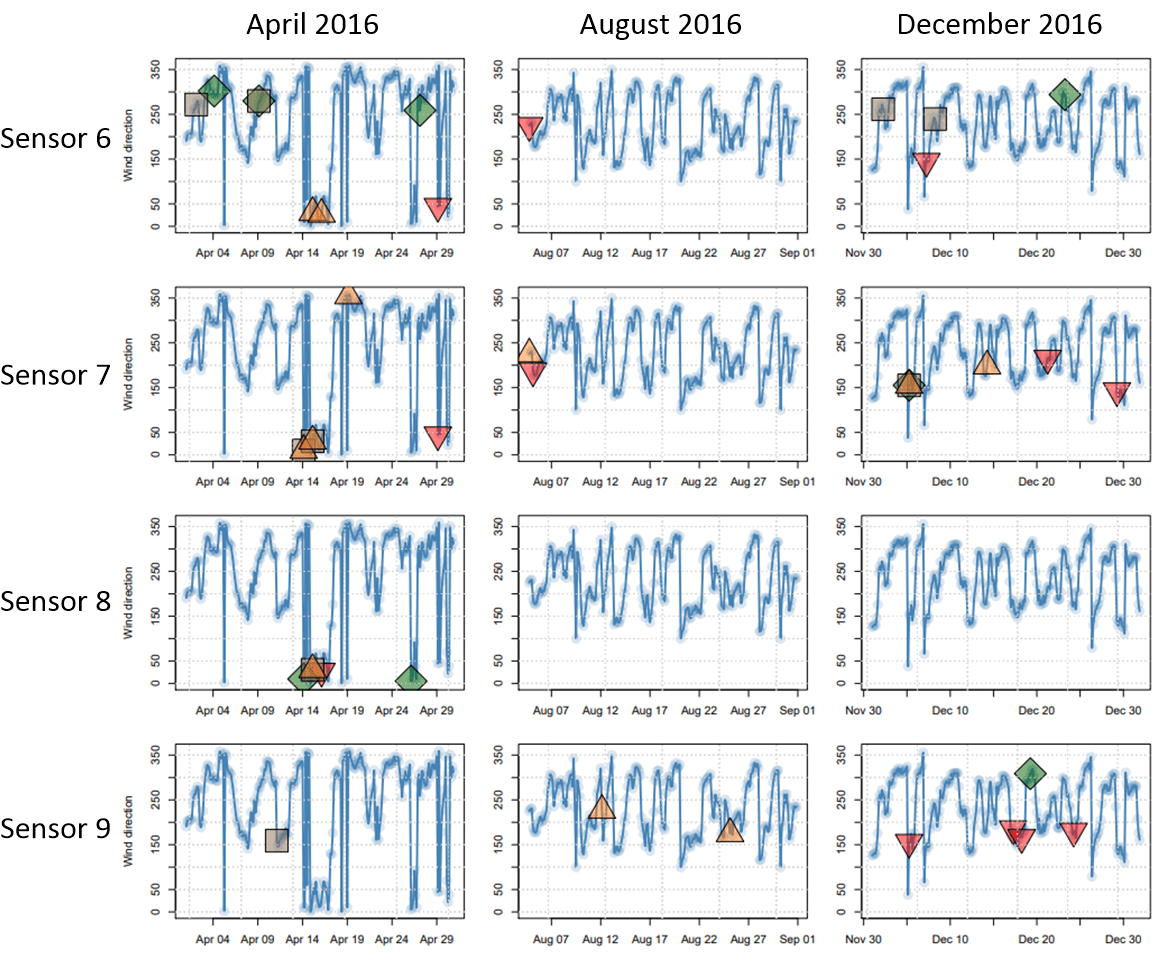}
\caption{Continue of Fig. \ref{figure_meteorological_matchedDates_timeSeries_1}.}
\label{figure_meteorological_matchedDates_timeSeries_2}
\end{figure*}

The information of detailed time within the month is missing in plots of Figures \ref{figure_meteorological_matchedDates_1} and \ref{figure_meteorological_matchedDates_2}. Therefore, it is also useful to have another set of plots where the time is also encoded. Figures \ref{figure_meteorological_matchedDates_timeSeries_1} and \ref{figure_meteorological_matchedDates_timeSeries_2} include the time series plots of the meteorological data where the rows an columns correspond to the nine sensors and the three months, respectively. The same color coding as in Figures \ref{figure_meteorological_matchedDates_1} and \ref{figure_meteorological_matchedDates_2} is utilized. The question is whether plotting the time series of wind direction or wind speed is better. For figuring out which factory was responsible for which chemical, the information of wind direction is more essential. Therefore, we plot the time series of wind direction. 

From Figures \ref{figure_meteorological_matchedDates_1}, \ref{figure_meteorological_matchedDates_2}, \ref{figure_meteorological_matchedDates_timeSeries_1}, and \ref{figure_meteorological_matchedDates_timeSeries_2}, we can infer which factories are responsible for the high values of chemicals. In order to analyze in a better way, we write the information obtained from Figures \ref{figure_meteorological_matchedDates_1}, \ref{figure_meteorological_matchedDates_2}, \ref{figure_meteorological_matchedDates_timeSeries_1}, and \ref{figure_meteorological_matchedDates_timeSeries_2} in Tables \ref{table_results_prediction_Appluimonia}, \ref{table_results_prediction_Chlorodinine}, \ref{table_results_prediction_Methylosmolene}, and \ref{table_results_prediction_AGOC3A} each of which for a chemical. For every chemical, we list the sensor, the date, and the wind direction in the tables. for more convenience and better visualization, the wind direction is shown by \textbf{arrow glyphs}.

Using the wind direction and the sensor index and by observing the location of sensors in Fig. \ref{figure_map}, we list the suspect factories which might have caused the high values of chemicals. Notice that some wind directions for some sensors do not give us any suspect or can accuse all the factories. In the tables, we write a dash and ``all'' for these two cases, respectively. 
As can be seen in Tables \ref{table_results_prediction_Appluimonia}, \ref{table_results_prediction_Chlorodinine}, \ref{table_results_prediction_Methylosmolene}, and \ref{table_results_prediction_AGOC3A}, \textbf{for different months and for almost all the four chemicals, the two factories Roadrunner and Kasios are suspect}. According to Table \ref{table_factories}, Roadrunner Fitness Electronics produces fitness trackers, heart rate monitors, and sport-related products and Kasios Office Furniture makes metal and composite-wood office furniture. It is interesting that the first glance at Table \ref{table_factories} makes us suspicious to Radiance ColourTek and Kasios Office Furniture because producing their products seem to be more harmful to environment. However, the analysis of data using data visualization gives us clue that the two factories Roadrunner Fitness Electronics and Kasios Office Furniture are suspect. \textbf{Kasios Office Furniture makes sense to make some trouble because of their metal products. The reason that Roadrunner Fitness Electronics might have made some trouble is probably because of some chemicals they are using for building their sport-related products}.

From Tables \ref{table_results_prediction_Appluimonia}, \ref{table_results_prediction_Chlorodinine}, \ref{table_results_prediction_Methylosmolene}, and \ref{table_results_prediction_AGOC3A}, we can also see that:
\begin{itemize}
\item \textbf{The high values of Appluimonia have often happened in August and December 2016.}
\item \textbf{The high values of Chlorodinine have often happened in April and December 2016.}
\item \textbf{The high values of Methylosmolene have often happened in April and December 2016.}
\item \textbf{The high values of AGOC-3A have often happened in all April, August, and December 2016.}
\end{itemize}

\section{Analysis of Aerial Images}

\subsection{The Multi-Channel Images}

There are $12$ multi-channel aerial images of the preserve provided in the data subset 3. The sizes of images are $650 \times 650$ pixels and they are taken from the preserve in different seasons of years 2014 to 2016. The channels of the images are blue, green, red, Near Infrared (NIR), Short-Wave Infrared (SWIR) 1, and Short-Wave Infrared (SWIR) 2. These channels include different bands in the electromagnetic spectrum. These bands are reported in Table \ref{table_bands} where their characteristics are also mentioned. 

\begin{table*}[!h]
\setlength\extrarowheight{5pt}
\centering
\scalebox{0.75}{    
\begin{tabular}{l | c | c | l}
\hline
\hline
\textbf{Band} & \textbf{Color} & \textbf{Wavelength (nm)} & \textbf{Useful for Mapping}\\
\hline
\hline
B1 & Blue & 450-520 & penetrates water, shows thin clouds\\
\hline
B2 & Green & 520-600 & shows different types of plants\\
\hline
B3 & Red & 630-690 & shows vegetation color and mineral deposits\\
\hline
B4 & NIR & 770-900 & partially absorbed by water, shows chlorophyll and vegetation\\
\hline
B5 & SWIR 1 & 1550-1750 & absorbed by liquid water, shows moisture of soil and vegetation\\
\hline
B6 & SWIR 2 & 2090-2350 & insenitive to vegetation, shows differences in soil mineral\\
\hline
\hline
\end{tabular}%
}
\caption{Different bands of multi-channel images and their characteristics.}
\label{table_bands}
\end{table*}

\subsection{Estimation of Scale and Orientation of Images}

\begin{figure}[!t]
\centering
\includegraphics[width=3.25in]{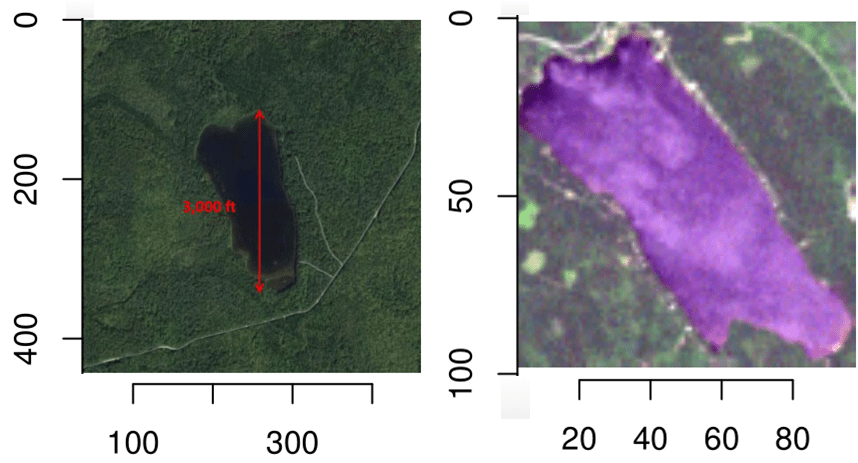}
\caption{The Boonsong lake within the preserve. Left: the provided RGB image of this lake in the dataset. Right: The cropped image of the lake from the RGB image of June 2016.}
\label{figure_boonsong_lake}
\end{figure}

In the dataset, there is an RGB image of the Boonsong lake, located within the preserve. This RGB image is shown in Fig. \ref{figure_boonsong_lake}. Furthermore, this lake was visually found in the RGB image of June 2016 and cropped from the image. The cropped image is also shown in Fig. \ref{figure_boonsong_lake}. The size of the cropped image is $(167-70+1) \times (420-325+1) = 98 \times 96$ pixels.
The length of this lake is given $3000$ feet. On the other hand, using Pythagorean theorem, we have:
\begin{align*}
& \ell = \sqrt{96^2 + 98^2} = 137.186 \text{ pixels} \overset{\text{set}}{=} 3000, \\
& \implies 1 \text{ pixel} = 21.86 \text{ feet} = 6.66 \text{ meters}.
\end{align*}
This gives us the scale and resolution of the provided multi-channel images:
\begin{align*}
650 \text{ pixels} \times 650 \text{ pixels} &= 14209 \text{ feet} \times 14209 \text{ feet}, \\
& = 4.329 \text{ km} \times 4.329 \text{ km}.
\end{align*}
The orientation of the images can also be found using the orientation of the image of Boonsong lake. This lake is oriented north-south; hence, the images are oriented from north-east at top to south-west at bottom.

\subsection{Analysis of RGB Images}

\begin{figure*}[!t]
\centering
\includegraphics[width=6in]{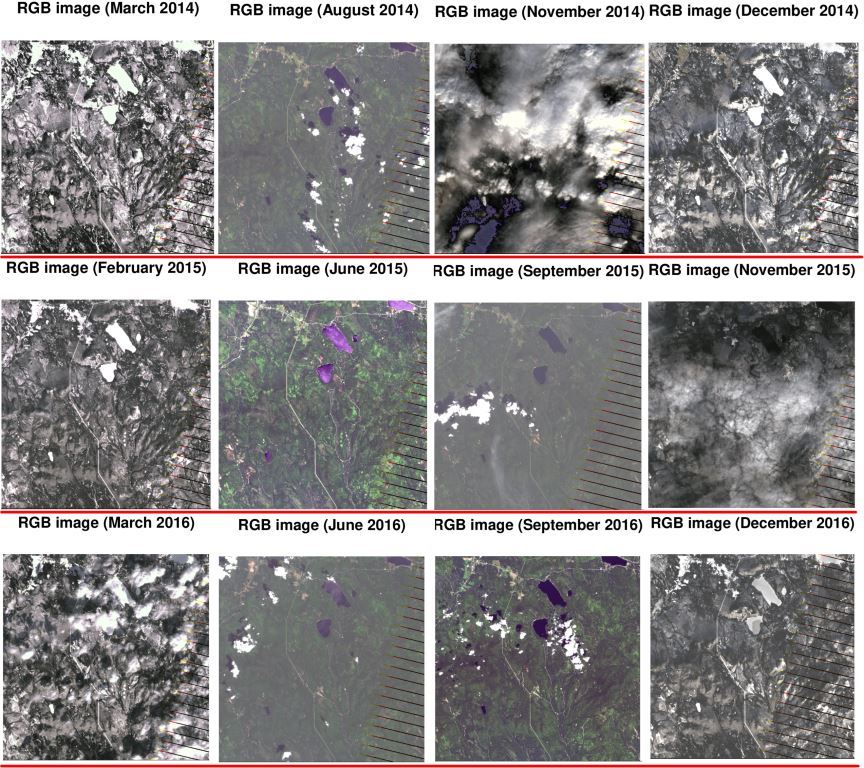}
\caption{The RGB aerial images from the preserve during years 2014 to 2016.}
\label{figure_RGB_images}
\end{figure*}

Using the three first channels, we can have the RGB images shown in Fig. \ref{figure_RGB_images}. As can be seen in this figure, for every year, we have four images with the order of late winter, summer, early fall, and late fall. The images of winter (first column) and late fall (last column) are showing snow, ice, and frozen lake as expected.

Two of the images which are images of November 2014 and November 2015 are covered by \textbf{clouds} completely. Some small clouds also exist in images of August 2014, September 2015, March 2016, June 2016, and September 2016.

From the RGB images, it can be seen that \textbf{in summer and early fall, we have more vegetation as expected. However, in winter and late fall, the vegetation declines slightly because of the cold}.

\subsection{Analysis of Changes In Vegetation Health}

\begin{figure*}[!t]
\centering
\includegraphics[width=6in]{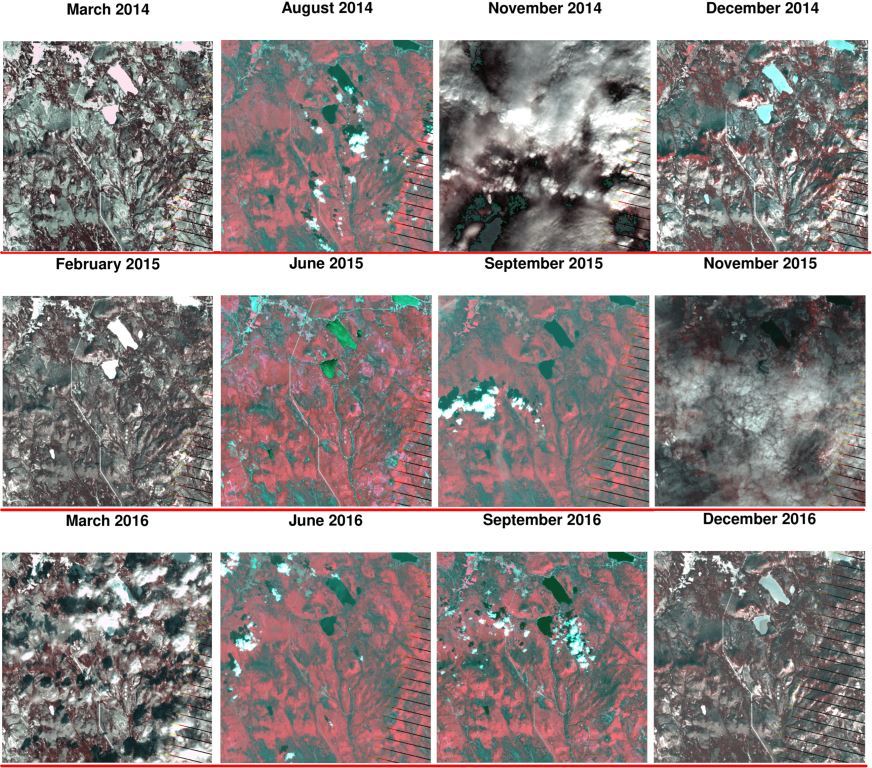}
\caption{The images obtained from bands [B4, B3, B2] from the multi-channel images of the preserve during years 2014 to 2016.}
\label{figure_false_images_1}
\end{figure*}

According to Table \ref{table_bands}, the bands B3 and B4 represent the vegetation and chlorophyll. Therefore, if we make \textbf{false-color images} using the three channels of \textbf{[B4, B3, B2]}, we can show the changes in plant health. These images are shown in Fig. \ref{figure_false_images_1}. The red colors in these false-color images show the good plant health.

\begin{figure*}[!t]
\centering
\includegraphics[width=6in]{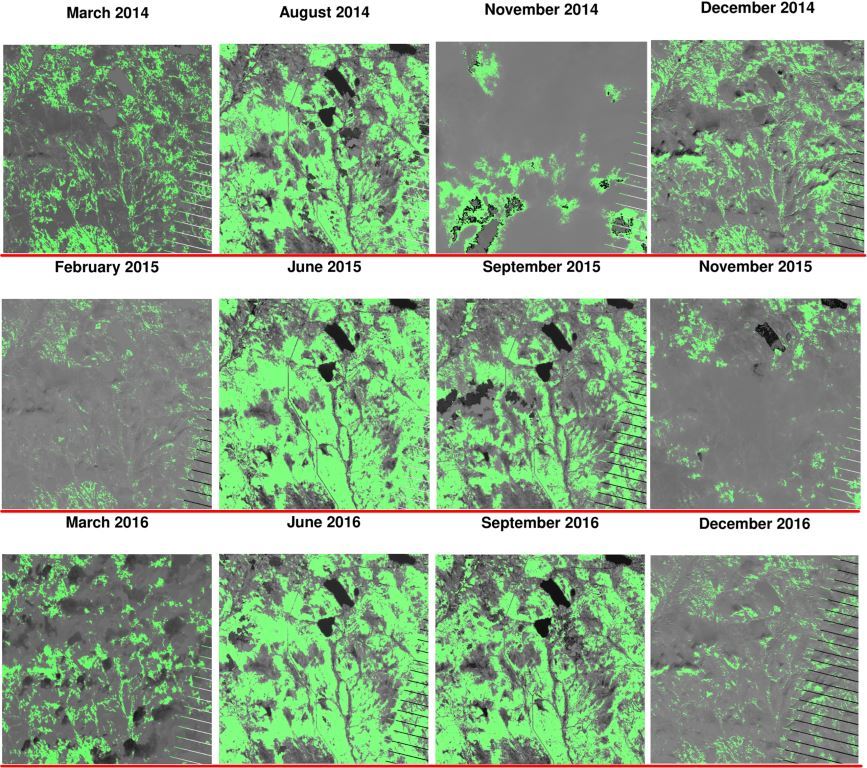}
\caption{The Normalized Difference Vegetation Index (NVDI) of the preserve during years 2014 to 2016. The threshold for highlighting the healthy vegeterated areas is 0.1 out of 1.}
\label{figure_NDVI}
\end{figure*}

Another way to assess the vegetation health is to plot the Normalized Difference Vegetation Index (NDVI). The NDVI can be calculated as this ratio \cite{web_NDVI}:
\begin{align*}
\text{NDVI} = \frac{\text{B4} - \text{B3}}{\text{B4} + \text{B3}}.
\end{align*}
This ratio contains both B3 and B4 bands which are sensitive to vegetation and plant health. In Fig. \ref{figure_NDVI}, we plot this ratio across the image frame in order to assess the vegetation. In this figure, the NDVI index above the threshold $0.1$ out of $1$ is highlighted by green color to show good health of vegetation.

Note that in Figures \ref{figure_false_images_1} and \ref{figure_NDVI}, images of November 2014 and November 2015 might not be valid for consideration because of existence of thick clouds. As can be seen in images of Figures \ref{figure_false_images_1} and \ref{figure_NDVI}:
\begin{itemize}
\item \textbf{In summer and early fall, the vegetation health is better} which is expected because of suitable weather conditions.
\item By noticing March 2014 and February 2015, the \textbf{vegetation of winter 2015 has declined compared to summer 2014.}
\item By noticing August 2014 and June 2015, the \textbf{vegetation of summer 2015 has improved compared to summer 2014.}
\item By noticing images of years 2015 and 2016, the \textbf{overall vegetation of 2016 has improved compared to 2015.}
\item By noticing images of June 2015 and June 2016, the \textbf{vegetation of up-left corner of image has suspiciously declined. According to orientation of images, the vegetation of north and north-west of preserve has declined.}
\item Noticing all images of 2014, 2015, and 2016, we see that \textbf{the overall vegetation of preserve is declining across the years.}
\end{itemize}

\subsection{Analysis of Dry Areas}

\begin{figure*}[!t]
\centering
\includegraphics[width=6in]{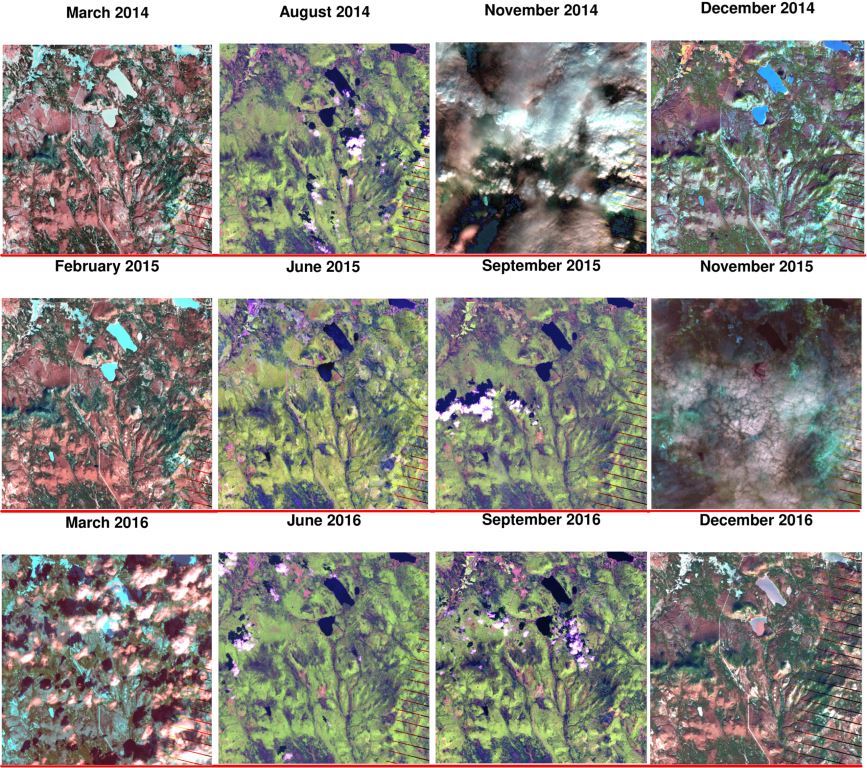}
\caption{The images obtained from bands [B5, B4, B2] from the multi-channel images of the preserve during years 2014 to 2016.}
\label{figure_false_images_2}
\end{figure*}

According to Table \ref{table_bands}, the bands B4 and B5 represent the dry areas because they are absorbed by liquid water. Therefore, if we make \textbf{false-color images} using the three channels of \textbf{[B5, B4, B2]}, we can show the dry regions of the preserve. These images are shown in Fig. \ref{figure_false_images_2}. The red colors in these false-color images show the dry or burned regions.

\begin{figure*}[!t]
\centering
\includegraphics[width=6in]{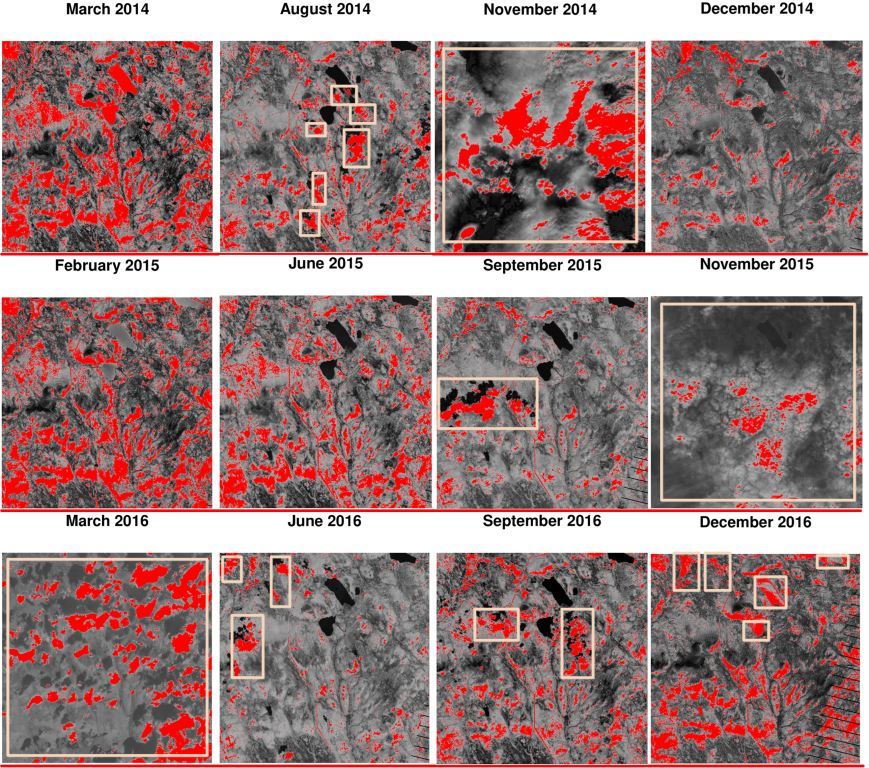}
\caption{The images obtained from the band B5 of the multi-channel images of the preserve during years 2014 to 2016. The threshold for highlighting the dry or burned areas is 0.7 out of 1. The areas which are falsely colored red because of clouds, lake, or ice are shown by rectangles.}
\label{figure_burn}
\end{figure*}

Another way to observe the dry regions of the preserve is to merely consider the \textbf{band B5} which is completely absorbed by liquid water (see Table \ref{table_bands}). In Fig. \ref{figure_burn}, we plot this band across the image frame in order to observe the dry regions. In this figure, the value of B5 above the threshold $0.7$ out of $1$ is highlighted by red color to show the dry regions. In the images of this figure, some areas are falsely highlighted because of the clouds. These areas are shown by rectangles to be excluded from our analysis.

As can be seen in figures \ref{figure_false_images_2} and \ref{figure_burn}: 
\begin{itemize}
\item Noticing images of August 2014 and June 2015, we can say that \textbf{2015 has been drier than 2014}.
\item Noticing images of June 2015 and June 2016, we can say that \textbf{2016 has had more moisture than 2015 in summer}.
\item Noticing images of September 2015 and September 2016, we can say that \textbf{2016 has been drier than 2015 in fall}.
\item Noticing all images of 2014, 2015, and 2016, we see that \textbf{overall, the preserve is becoming drier and drier across the years.}
\end{itemize}

\subsection{Analysis of Soil Mineral Content}

\begin{figure*}[!t]
\centering
\includegraphics[width=6in]{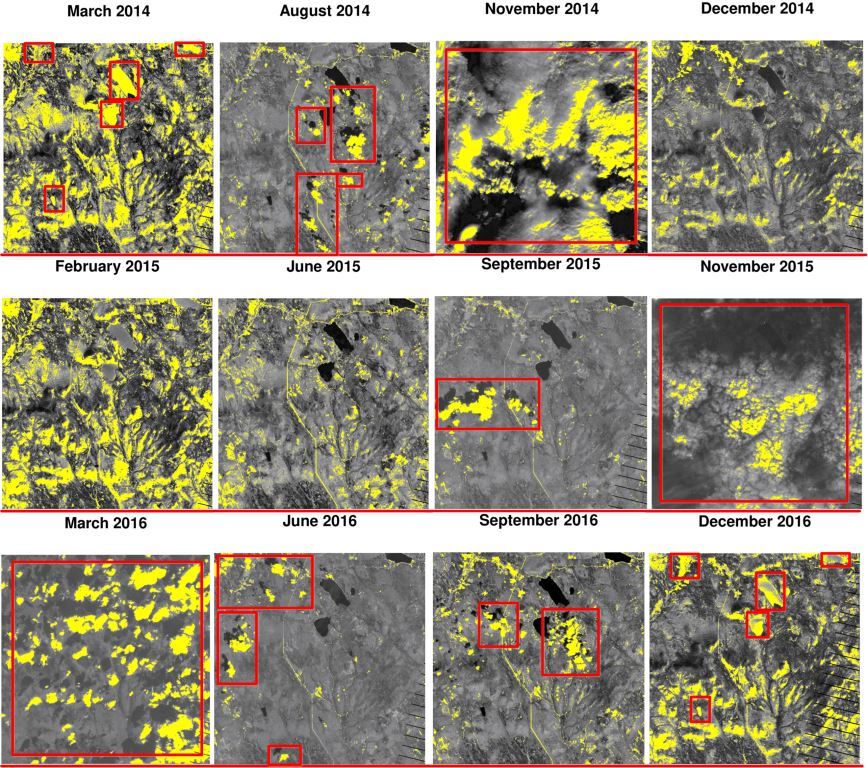}
\caption{The images obtained from the band B6 of the multi-channel images of the preserve during years 2014 to 2016. The threshold for highlighting the rich areas in minerals is 0.7 out of 1. The areas which are falsely colored yellow because of clouds, lake, or ice are shown by rectangles.}
\label{figure_mineral}
\end{figure*}

According to Table \ref{table_bands}, the \textbf{band B6} is sensitive to the soil mineral content. In Fig. \ref{figure_mineral}, we plot this band across the image frame in order to observe the mineral content of soil. In this figure, the value of B6 above the threshold $0.7$ out of $1$ is highlighted by yellow color to show the rich mineral amount of soil. Note that the mineral of soil is useful for plant to grow. In the images of this figure, some areas are falsely highlighted because of the clouds. These areas are shown by rectangles to be excluded from our analysis.

As can be seen in Fig. \ref{figure_mineral}: 
\begin{itemize}
\item Noticing images of August 2014 and June 2015, we can say that \textbf{2016 has been slightly richer in soil mineral than 2015 in summer}.
\item Noticing images of June 2015 and June 2016, we can say that \textbf{2016 has been poorer in soil mineral than 2015 in summer}.
\item Noticing images of September 2015 and September 2016, we can say that \textbf{2016 has been slightly richer in soil mineral than 2015 in fall}.
\item Noticing all images of 2014, 2015, and 2016, we see that \textbf{overall, the preserve is becoming slightly richer in soil mineral.}
\end{itemize}

If we combine the observation of Figures \ref{figure_NDVI} and \ref{figure_mineral}, we see that \textbf{the vegetation is declining but soil mineral is improving across the years}. This shows that \textbf{the reason of vegetation decline is not because of the soil but it must be the chemicals in the air. This gives us some clues back to the previous analysis of the chemicals in last section.}

\section{Conclusion}\label{section_conclusions}

\subsection{Summary of The Findings}

In the previous sections, we mentioned various large and small suspicious activities and events which may have happened in the preserve. Here, we summarize the most important findings and do not re-mention the small details or not very important conclusions.

The main important conclusions from the \textbf{analysis of vehicle traffic} through the nature preserve were:
\begin{itemize}
\item The data of some two-axle cars, two-axle trucks, three-axle trucks, and two-axle buses have not been recorded after some camping and general gates.
\item Some four-axle trucks have passed through the gates illegally. These gates are supposed to be passed only by rangers.
\item Rangers have missed inspecting in the east of the preserve mostly. However, non-ranger vehicles have visited the east significantly.
\item Around June 2015, the traffic of all types of vehicles has increased a lot. This might be destructive to the nature preserve or might have had bad effects on the quality of life of flora and fauna in the preserve.
\end{itemize}

The main important conclusions from the \textbf{analysis of surrounding factories}:
\begin{itemize}
\item Some sensor failures have occurred mostly at three time periods, i.e., April and June 2016, August and September 2016, and December 2016.
\item At some periods of time, we have had very high values (30 to 80 parts per million) of Methylosmolene and AGOC-3A. These two chemicals are very dangerous and toxic so they are very destructive to the life of flora and fauna in the preserve.
\item Some odd chemical accumulation of Appluimonia and Chlorodinine have happened at different months of 2016 recorded by sensor 6. The values of these two chemicals have increased in 2016. The sensor 6 is close to all the factories.
\item In different months of 2016, the two factories Roadrunner and Kasios are suspect to generating the four chemicals Appluimonia, Chlorodinine, Methylosmolene, and AGOC-3A higher than the acceptable standards.
\end{itemize}

The main important conclusions from the \textbf{analysis of aerial multi-channel images}:
\begin{itemize}
\item Overall, the nature preserve is becoming drier and drier across the years 2014 to 2016.
\item Overall, the nature preserve is becoming poorer in vegetation across the years 2014 to 2016.
\item The vegetation of the preserve is declining but the soil mineral is improving across the years. This shows that the reason of vegetation decline is not because of the soil but it must be the chemicals in the air.
\end{itemize}

\subsection{Combining The Findings}

If we see the different findings mentioned in the previous section, we can come up with some combined findings which might be useful for final analysis:
\begin{itemize}
\item The significant number of visits to the preserve in June 2015 might have had destructive influence on the preserve resulting in drier soil and decline in vegetation.
\item The increase of toxic chemicals generated by the two factories Roadrunner and Kasios in 2016 might have had bad impact on the breath of plants resulting in decline of vegetation. 
\item The increase of toxic chemicals might also have had destructive effect on the life of animals and birds including the Blue Pipit.
\item The odd accumulation of the chemicals recorded by sensor 6 and the strange failures in the performance of sensors might have some relations.
\end{itemize}

\subsection{The Possible Reasons of Population Decline of The Blue Pipit}

Finally, we can list the possible reasons of the decline in population of the Blue Pipit:
\begin{itemize}
\item The high number of visits to the preserve in June 2015.
\item Some strange activities in the east of the preserve while the inspection of rangers was not satisfactory there.
\item Some illegal passes through the gates. The destructive activities might have happened in those cases.
\item Some strange ending points of trajectories for some vehicles. The destructive activities might have happened in those cases.
\item Some odd problems in performance of the sensors at the south of the preserve in different months of 2016. The sensors might have been manipulated for evil reasons by either the visitors or the factory managers.
\item Generation of toxic chemicals by the two factories Roadrunner and Kasios in different months of 2016.
\item The soil of the preserve has become dry and the vegetation health has declined.
\end{itemize}

\subsection{Suggested Solutions For Preventing Further Problems}

As a conclusion to the paper, we suggest some solutions for preventing similar problems in the preserve:
\begin{itemize}
\item Controlling the number of visitors.
\item Permanent investigation in ``all'' regions of the preserve and in ``various times''.
\item Increasing security at the gates.
\item Controlling whether the visitors have left the preserve or not after a reasonable time.
\item Controlling and investigating the chemicals generated by the factories more carefully.
\item Using fertilizers regularly in order to make the soil of the preserve richer for better vegetation. Better vegetation and flora helps to have better life for fauna.
\end{itemize}

\section*{Acknowledgment}
The authors hugely thank Prof. Wayne Oldford, professor of Statistics and Actuarial Science at the University of Waterloo, for his helpful guides, fruitful discussions, and his course ``Data Visualization''. 


\bibliography{References}

\begin{thebibliography}{6}
\providecommand{\natexlab}[1]{#1}
\providecommand{\url}[1]{\texttt{#1}}
\expandafter\ifx\csname urlstyle\endcsname\relax
  \providecommand{\doi}[1]{doi: #1}\else
  \providecommand{\doi}{doi: \begingroup \urlstyle{rm}\Url}\fi

\bibitem[Barnett(1974)]{barnett1974elements}
Barnett, Vick.
\newblock \emph{Elements of sampling theory}.
\newblock English Universities Press, London, 1974.

\bibitem[Friedman et~al.(2009)Friedman, Hastie, and
  Tibshirani]{friedman2001elements}
Friedman, Jerome, Hastie, Trevor, and Tibshirani, Robert.
\newblock \emph{The elements of statistical learning}, volume~2.
\newblock Springer series in statistics New York, NY, USA:, 2009.

\bibitem[{NASA Earth Observatory}(2000)]{web_NDVI}
{NASA Earth Observatory}.
\newblock Measuring vegetation.
\newblock
  \url{https://earthobservatory.nasa.gov/features/MeasuringVegetation/measuring_vegetation_2.php},
  2000.
\newblock Accessed: 2018-06-10.

\bibitem[Rutstrum(2000)]{rutstrum2000wilderness}
Rutstrum, Calvin.
\newblock \emph{The wilderness route finder: the classic guide to finding your
  way in the wild}.
\newblock University of Minnesota Press, 2000.

\bibitem[{Visual Analytics Community}(2017)]{web_vast_challenge}
{Visual Analytics Community}.
\newblock Vast challenge 2017.
\newblock \url{http://www.vacommunity.org/VAST+Challenge+2017}, 2017.
\newblock Accessed: 2018-06-10.

\bibitem[Waddell \& Oldford(2018)Waddell and Oldford]{web_loon}
Waddell, Adrian and Oldford, Wayne.
\newblock Loon: An interactive statistical visualization toolkit.
\newblock \url{http://waddella.github.io/loon/}, 2018.
\newblock Accessed: 2018-06-10.

\end{thebibliography}
\bibliographystyle{icml2016}

\end{document}